\documentclass{article}

\usepackage{microtype}
\usepackage{graphicx}
\usepackage{booktabs} 

\usepackage{times}
\usepackage{url}

\usepackage{color}
\usepackage{amsmath,amssymb,amsthm,bm,bbm}

\usepackage{multirow}
\usepackage{multicol}
\usepackage{graphicx}

\usepackage{wrapfig}
\usepackage{lipsum}  

\usepackage{algorithm}
\usepackage{algorithmic}

\usepackage{subcaption}
\usepackage{graphicx}

\usepackage{epstopdf}
\usepackage{enumerate}

\usepackage{float}
\usepackage{tabularx}
\usepackage{makecell}
\usepackage{arydshln}

\usepackage{enumitem}

\usepackage{anyfontsize}
\usepackage{makecell}

\usepackage{hyperref}

\newcommand{\xv}{{\mathbf x}}

\newcommand{\zv}{{\mathbf z}}

\newcommand{\ev}{{\mathbf e}}

\newcommand{\Xv}{{\mathbf X}}

\newcommand{\Cc}{{\mathcal C}}

\newcommand{\Ec}{{\mathcal E}}

\newcommand{\Kc}{{\mathcal K}}

\newcommand{\Tc}{{\mathcal T}}

\newcommand{\Xc}{{\mathcal X}}
\newcommand{\Yc}{{\mathcal Y}}
\newcommand{\Zc}{{\mathcal Z}}

\newcommand{\loss}{{\mathcal L}}
\newcommand{\data}{{\mathcal D}}

\newcommand{\E}{{\mathbb{E}}}

\newcommand{\proposed}{{AC-TPC}}

\newcolumntype{L}[1]{>{\raggedright\let\newline\\\arraybackslash\hspace{0pt}}m{#1}}
\newcolumntype{C}[1]{>{\centering\let\newline\\\arraybackslash\hspace{0pt}}m{#1}}
\newcolumntype{R}[1]{>{\raggedleft\let\newline\\\arraybackslash\hspace{0pt}}m{#1}}

\newsavebox\CBox

\usepackage[accepted]{icml2020}

\icmltitlerunning{Temporal Phenotyping using Deep Predictive Clustering of Disease Progression}

\begin{document}

\twocolumn[
\icmltitle{Temporal Phenotyping using Deep Predictive Clustering of Disease Progression}



\icmlsetsymbol{equal}{*}

\begin{icmlauthorlist}
	\icmlauthor{Changhee Lee}{1}
	\icmlauthor{Mihaela van der Schaar}{2,3,1}
\end{icmlauthorlist}

\icmlaffiliation{1}{University of California, Los Angeles,  USA}
\icmlaffiliation{2}{University of Cambridge, UK}
\icmlaffiliation{3}{Alan Turing Institute, UK}
\icmlcorrespondingauthor{Changhee Lee}{chl8856@ucla.edu}

\icmlkeywords{Machine Learning, ICML}

\vskip 0.3in
]



\printAffiliationsAndNotice{}  

\begin{abstract}
	Due to the wider availability of modern electronic health records, patient care data is often being stored in the form of time-series. 
	Clustering such time-series data is crucial for patient phenotyping, anticipating patients' prognoses by identifying ``similar'' patients, and designing treatment guidelines that are tailored to homogeneous patient subgroups. 
	In this paper, we develop a deep learning approach for clustering time-series data, where each cluster comprises patients who share similar future outcomes of interest (e.g., adverse events, the onset of comorbidities). 	
	To encourage each cluster to have homogeneous future outcomes, the clustering is carried out by learning discrete representations that best describe the future outcome distribution based on novel loss functions.
	Experiments on two real-world datasets show that our model achieves superior clustering performance over state-of-the-art benchmarks and identifies meaningful clusters that can be translated into actionable information for clinical decision-making.
\end{abstract}

\section{Introduction}
Chronic diseases -- such as cystic fibrosis and dementia -- are heterogeneous in nature, with widely differing outcomes even in narrow patient subgroups. 
Disease progression manifests through a broad spectrum of clinical factors, collected as a sequence of measurements in electronic health records, which gives a rise to complex progression patterns among patients \citep{Samal:11,Jinsung:JBHI17}. 
For example, cystic fibrosis evolves slowly, allowing for development of comorbidities and bacterial infections, and creating distinct responses to therapeutic interventions, which in turn makes the survival and quality of life substantially different \citep{CF_ref:12,Changhee:TBME19}.
Identifying patient subgroups with similar progression patterns can be advantageous for understanding such heterogeneous diseases. This allows clinicians to anticipate patients' prognoses by comparing to ``similar'' patients and to design treatment guidelines tailored to homogeneous subgroups \citep{FeiWang:18}. 

\begin{figure}[t!]
	\centering 
	\includegraphics[width=3.3in, trim= 0.1 0.1 0.1 0.1]{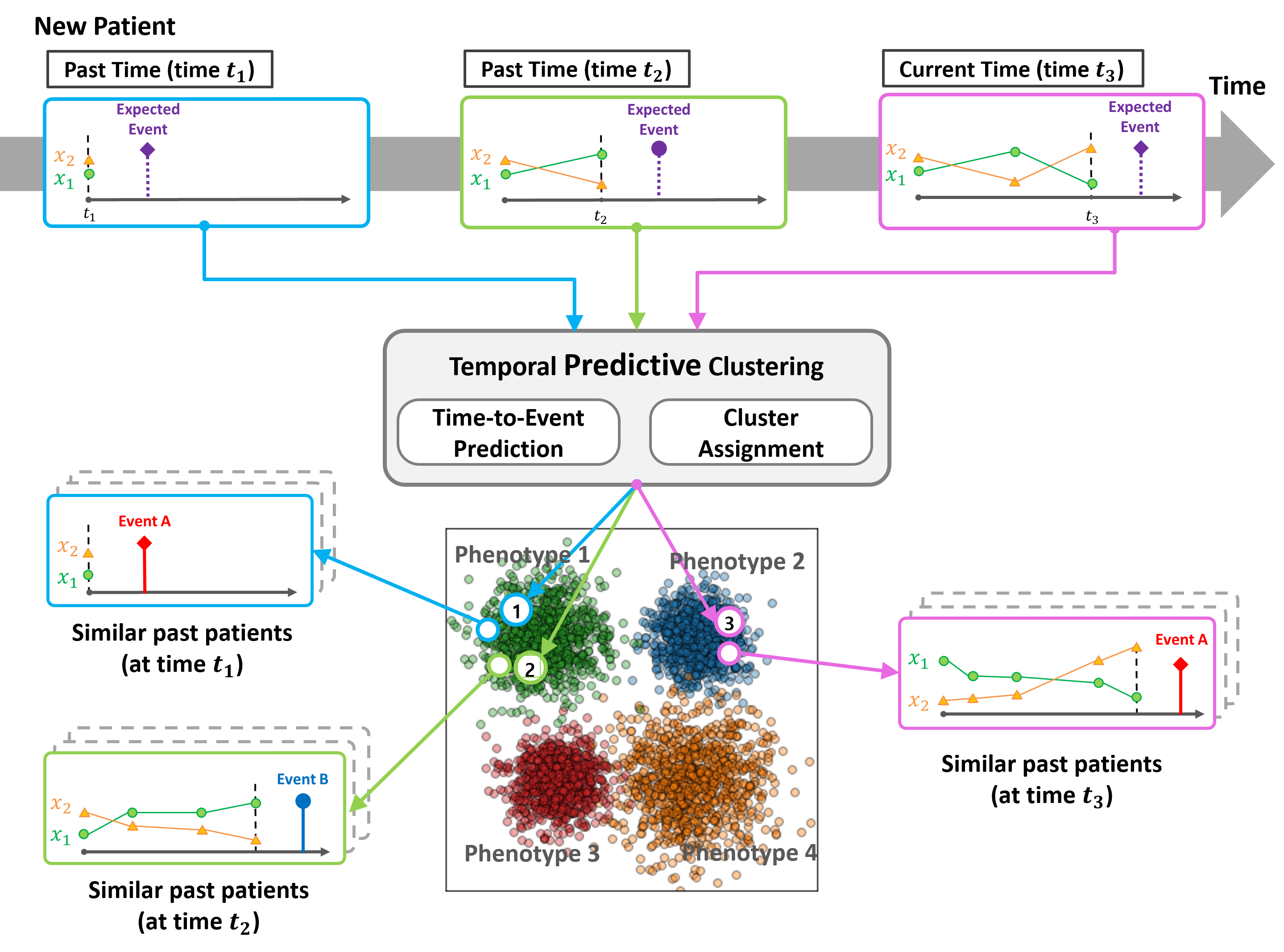} \vspace{-6.5mm}
	\caption{A conceptual illustration of our (real-time) clustering procedure. Here, a new patient is assigned over time to one of the four phenotypes based on the expected future event -- either Event A or Event B -- as new observations are collected.} \label{fig:concept} \vspace{-4.5mm}
\end{figure}
Temporal clustering has been recently used as a data-driven framework to partition patients with time-series observations into subgroups of patients.
Recent research has typically focused on either finding fixed-length and low-dimensional representations \citep{FeiWang:18,Rusanov:16} or on modifying the similarity measure \citep{Giannoula:18,Luong:17} both in an attempt to apply the existing clustering algorithms to time-series observations.
However, clusters identified from these approaches are purely unsupervised -- they do not account for patients' observed outcomes (e.g., adverse events, the onset of comorbidities, etc.) -- which leads to heterogeneous clusters if the clinical presentation of the disease differs even for patients with the same outcomes.
Thus, a common prognosis in each cluster remains unknown which can mystify the understanding of the underlying disease progression \citep{Boudier:19,comorbidity_ref:1}. 
To overcome this limitation, we focus on \textit{predictive clustering} \citep{Blockeel:17} to combine predictions on the future outcomes with clustering.
More specifically, we aim at finding cluster assignments and centroids by learning discrete representations of time-series that best describe the future outcome distribution. By doing so, patients in the same cluster share similar future outcomes to provide a prognostic value. 
Figure \ref{fig:concept} illustrates a pictorial depiction of the clustering procedure. 

In this paper, we propose an actor-critic approach for temporal predictive clustering, which we call \proposed.\footnote{Source code available at \url{https://github.com/chl8856/AC_TPC}.}  
Our model consists of three networks -- an \textit{encoder}, a \textit{selector}, and a \textit{predictor} -- and a set of centroid candidates. 
The key insight, here, is that we model temporal predictive clustering as learning discrete representations of the input time-series that best describe the future outcome distribution. More specifically, the encoder maps an input time-series into a continuous latent encoding; the selector assigns a cluster (i.e., maps to a discrete representation) to which the input belongs by taking the latent encoding as an input; and the predictor estimates the future outcome distributions conditioned on either the encoding or the centroid of the selected cluster (i.e., the selected discrete representation). 
The following three contributions render our model to achieve our goal. First, to encourage homogeneous future outcomes in each cluster, we define a clustering objective based on the Kullback-Leibler (KL) divergence between the predictor's output given the time-series, and that given the assigned centroids. Second, we transform solving a combinatorial problem of identifying clusters into iteratively solving two sub-problems: optimization of the cluster assignments and optimization of the centroids.  Finally, we allow ``back-propagation'' through the sampling process of the selector by adopting actor-critic training \citep{actor_critic:00}.


Throughout the experiments, we show significant performance improvements over the state-of-the-art clustering methods on two real-world medical datasets.
To demonstrate the practical significance of our model, we consider a more realistic scenario where the future outcomes of interest are high-dimensional -- that is, development of multiple comorbidities in the next year -- and interpreting all possible combinations is intractable.
Our experiments show that our model can identify meaningful clusters that can be translated into actionable information for clinical decision-making.

\section{Problem Formulation}
Let $\Xv \in \Xc$ and $Y \in \Yc$ be random variables for an input feature and an output label (i.e., one or a combination of future outcome(s) of interest) with a joint distribution $p_{XY}$ (and marginal distributions are $p_{X}$ and $p_{Y}$) where $\Xc$ is the feature space and $\Yc$ is the label space. Here, we focus our description on $C$-class classification tasks, i.e., $\Yc = \{1, \cdots, C\}$.\footnote{In the Supplementary Material, we discuss simple modifications for regression $\Yc = \mathbb{R}$ and $M$-dimensional binary classification tasks $\Yc = \{0,1\}^{M}$.}
We are given a time-series dataset $\data = \{ (\xv_{t}^{n}, y_{t}^{n})_{t=1}^{T^{n}} \}_{n=1}^{N}$ comprising sequences of realizations (i.e., observations) of the pair $(\Xv, Y)$ for $N$ patients.
Here, $(\xv_{t}^{n}, y_{t}^{n})_{t=1}^{T^{n}}$ is a sequence of $T^{n}$ observation pairs that correspond to patient $n$ and $t \in \Tc^{n} \triangleq \{1,\cdots,T^{n}\}$ denotes the time stamp at which the observations are made. 
From this point forward, we omit the dependency on $n$ when it is clear in the context and denote $\xv_{1:t} = (\xv_{1}, \cdots, \xv_{t})$. 

Our aim is to identify a set of $K$ \textit{predictive clusters}, $\Cc = \{ \Cc(1), \cdots, \Cc(K) \}$, for time-series data. 
Each cluster consists of homogeneous data samples, that can be represented by its centroid, based on a certain similarity measure.
There are two main distinctions from the conventional notion of clustering. 
First, we treat subsequences of each times-series as data samples and focus on partitioning $\{\{\xv_{1:t}^{n}\}_{t=1}^{T^{n}}\}_{n=1}^{N}$ into $\Cc$. 
Hence, we define a cluster as $\Cc(k) = \{\xv_{1:t}^{n}|t\in\Tc^{n},~s^{n}_{t} = k\}$ for $k\in \Kc \triangleq \{1,\cdots,K\}$ where $s_{t}^{n} \in \Kc$ is the cluster assignment for a given $\xv_{1:t}^{n}$.
This is to flexibly update the cluster assignment (in real-time) to which a patient belongs as new observations are being accrued over time.
Second, we define the similarity measure with respect to the label distribution and associate it with clusters to provide a prognostic value.  
More specifically, we want the distribution of output label for subsequences in each cluster to be homogeneous and, thus, can be well-represented by the centroid of that cluster. 

Let $S$ be a random variable for the cluster assignment -- that depends on a given subsequence $\xv_{1:t}$ -- and $Y|S=k$ be a random variable for the output given cluster $k$.
Then, such property of predictive clustering can be achieved by minimizing the following Kullback-Leibler (KL) divergence: $KL(Y_{t}|\Xv_{1:t}=\xv_{1:t} \| Y_{t}|S_{t}=k )$ for $\xv_{1:t} \in \Cc(k)$ which is defined as $\int_{y} p(y|\xv_{1:t}) \big(\log p(y|\xv_{1:t})   -\log p(y|s_{t}) \big)dy$ where $p(y|\xv_{1:t})$ and $p(y|s_{t})$ are the label distributions conditioned on a subsequence $\xv_{1:t}$ and a cluster assignment $s_{t}$, respectively.
Note that the KL divergence achieves its minimum when the two distributions are equivalent. 

Finally, we establish our goal as identifying a set of predictive clusters $\Cc$ that optimizes the following objective:
\begin{equation}  \label{eq:predictive_cluster_objective}
\underset{\Cc}{\text{minimize}} \! \sum_{k\in\Kc} \!\!\!\! \sum_{~~~\xv_{1:t}\in\Cc(k)} \!\!\!\!\!\!\!\! KL\big(Y_{t}|\Xv_{1:t}=\xv_{1:t} \big\| Y_{t}|S_{t}=k \big).
\end{equation}
Unfortunately, the optimization problem in \eqref{eq:predictive_cluster_objective} is highly non-trivial. 
We need to estimate the objective function in \eqref{eq:predictive_cluster_objective} while solving a non-convex combinatorial problem of finding the optimal cluster assignments and cluster centroids.

\section{Method: \proposed} \label{sec:method}
\begin{figure}[t!]
	\centering 
	\includegraphics[width=3.2in, trim= 0.1 0.1 0.1 0.1]{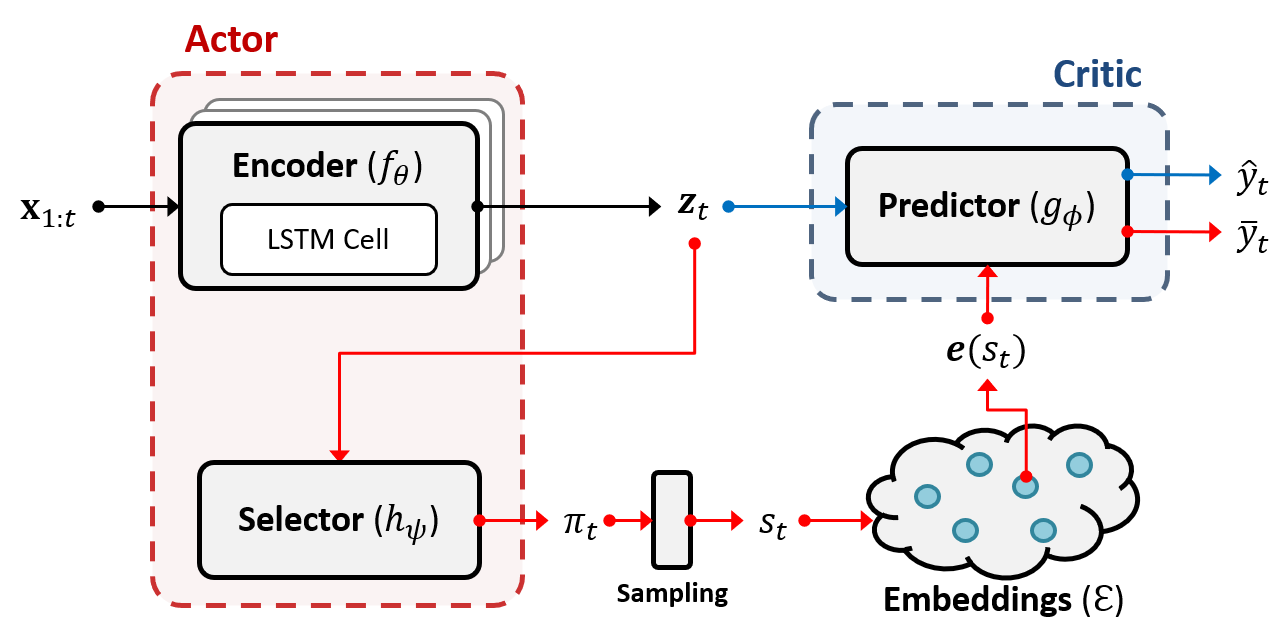}
	\caption{The block diagram of \proposed. The red line implies the procedure of estimating $p(y|S_{t}=s_{t})$ via a sampling process and the blue line implies that of estimating $p(y|\Xv_{1:t}=\xv_{1:t})$.} \label{fig:network_architecture} \vspace{-2mm}
\end{figure}
To effectively estimate the objective function in \eqref{eq:predictive_cluster_objective}, we introduce three networks -- an \textit{encoder}, a \textit{selector}, and a \textit{predictor} -- and an \textit{embedding dictionary} as illustrated in Figure \ref{fig:network_architecture}. 
These components together provide the cluster assignment and the corresponding centroid based on a given sequence of observations and enable us to estimate the probability density $p(y|s_{t})$. 
More specifically, we define each component as follows: \vspace{-2mm}
\begin{itemize}[leftmargin=1.2em]
	\item The \textit{encoder}, $f_{\theta}: \prod_{i=1}^{t} \Xc \rightarrow \Zc$, is a RNN (parameterized by $\theta$) that maps a (sub)sequence of a time-series $\xv_{1:t}$ to a latent representation (i.e., encoding) $\zv_{t} \in \Zc$ where $\Zc$ is the latent space. \vspace{-2mm}
	\item The \textit{selector}, $h_{\psi}: \Zc \rightarrow \Delta^{K-1}$, is a fully-connected network (parameterized by $\psi$) that provides a probabilistic mapping to a categorical distribution from which the cluster assignment $s_{t} \in \Kc$ is being sampled. \vspace{-2mm}
	\item The \textit{predictor}, $g_{\phi}: \Zc \rightarrow \Delta^{C-1}$, is a fully-connected network (parameterized by $\phi$) that estimates the label distribution given the encoding of a time-series or the centroid of a cluster. \vspace{-2mm}
	\item The \textit{embedding dictionary}, $\Ec = \{\ev(1),\cdots, \ev(K)\}$ where $\ev(k) \in \Zc$ for $k \in \Kc$, is a set of cluster centroids lying in the latent space which represents the corresponding cluster.
\end{itemize} \vspace{-3mm}
Here, $\Delta^{D-1} = \{ \mathbf{q} \in [0,1]^{D}: q_{1} + \cdots + q_{D} =1 \}$ is a $(D-1)$-simplex that denotes the probability distribution for a $D$-dimensional categorical (class) variable. 

At each time stamp $t$, the \textit{encoder} maps a input (sub)sequence $\xv_{1:t}$ into a latent encoding $\zv_{t} \triangleq f_{\theta}(\xv_{1:t})$.
Then, based on the encoding $\zv_{t}$, the cluster assignment $s_{t}$ is drawn from a categorical distribution that is defined by the \textit{selector} output, i.e., $s_{t}\sim Cat(\pi_{t})$ where $\pi_{t}= [\pi_{t}(1), \cdots, \pi_{t}(K)] \triangleq h_{\psi}(\zv_{t})$.
Once the assignment $s_{t}$ is chosen, we allocate the latent encoding $\zv_{t}$ to an embedding $\ev(s_{t})$ in the \textit{embedding dictionary} $\Ec$.
Since the allocated embedding $\ev(s_{t})$ corresponds to the centroid of the cluster to which $\xv_{1:t}$ belongs, we can, finally, estimate the density $p(y|s_{t})$ in \eqref{eq:predictive_cluster_objective} as the output of the \textit{predictor} given the embedding $\ev(s_{t})$, i.e., $\bar{y}_{t} \triangleq g_{\phi}(\ev(s_{t}))$.

\subsection{Loss Functions} \label{subsec:loss}
In this subsection, we define loss functions to achieve our objective in \eqref{eq:predictive_cluster_objective}; the details of how we train our model will be discussed in the following subsection.

\textbf{Predictive Clustering Loss: }
Since finding the cluster assignment of a given sequence is a probabilistic problem due to the sampling process, the objective function in \eqref{eq:predictive_cluster_objective} must be defined as an expectation over the cluster assignment.
Thus, we can estimate solving the objective problem in \eqref{eq:predictive_cluster_objective} as minimizing the following loss function:
\begin{equation} \label{eq:loss_predictive_clustering}
\loss_{1}(\theta, \psi, \phi, \Ec) = \E_{\xv,y\sim p_{XY}} \Big[ \sum_{t\in\Tc} \E_{s_{t} \sim Cat(\pi_{t})} \big[ \ell_{1}(y_{t}, \bar{y}_{t}) \big]\Big]
\end{equation}
where $\ell_{1}(y_{t}, \bar{y}_{t}) = - \sum_{c=1}^{C} y_{t}^{c} \log \bar{y}_{t}^{c}$. Here, we slightly abuse the notation and denote $y = [y^{1} \cdots y^{C}]$ as the one-hot encoding of $y$, and $y^{c}$ and $\bar{y}^{c}$ indicates the $c$-th component of $y$ and $\bar{y}$, respectively.
It is worth to highlight that minimizing $\ell_{1}$ is equivalent to minimizing the KL divergence in \eqref{eq:predictive_cluster_objective} since the former term of the KL divergence is independent of our optimization procedure.

One critical question that may arise is how to avoid trivial solutions in this unsupervised setting of identifying the cluster assignments and the centroids \citep{Yang:17}.
For example, all the embeddings in $\Ec$ may collapse into a single point or the selector simply assigns equal probability to all the clusters regardless of the input sequence.
In both cases, our model will fail to correctly estimate $p(y|s_{t})$ and, thus, end up finding a trivial solution. 
To address this issue, we introduce two auxiliary loss functions that are tailored to address this concern.
It is worth to highlight that these loss functions are not subject to the sampling process and their gradients can be simply back-propagated.

\textbf{Sample-Wise Entropy of Cluster Assignment: }
To motivate sparse cluster assignment such that the selector ultimately selects one dominant cluster for each sequence, we introduce sample-wise entropy of cluster assignment which is given as
\begin{equation} \label{eq:loss_sw_entropy}
\loss_{2}(\theta, \psi) = \E_{\xv \sim p_{X}} \Big[ - \sum_{t\in\Tc}\sum_{k \in \Kc} \pi_{t}(k) \log \pi_{t}(k) \Big]
\end{equation}
where $\pi_{t} =[\pi_{t}(1) \cdots \pi_{t}(K)] = h_{\psi}(f_{\theta}(\xv_{1:t}))$. The sample-wise entropy achieves its minimum when $\pi_{t}$ becomes an one-hot vector.

\textbf{Embedding Separation Loss: }
To prevent the embeddings in $\Ec$ from collapsing into a single point, we define a loss function that encourages the embeddings to represent different label distributions, i.e., $g_{\phi}(\ev(k))$ for $k\in \Kc$, from each other:
\begin{equation} \label{eq:loss_embedding_separation}
\loss_{3}(\Ec) = - \sum_{k\neq k'} \ell_{1}(g_{\phi}(\ev(k)), g_{\phi}(\ev(k')))
\end{equation}
where $\ell_{1}$ is reused to quantify the distance between label distributions conditioned on each cluster. 
We minimize \eqref{eq:loss_embedding_separation} when updating the embedding vectors $\ev(1), \cdots, \ev(K)$.

\begin{figure*} [!htp]
	\normalsize
	\begin{equation} \label{eq:loss_1_derivative}
	\nabla_{\!\omega_{A}} \loss_{A}(\theta, \psi, \phi) 
	= \E_{\xv,y \sim p_{XY}} \!\Big[ \sum_{t\in\Tc} \E_{s_{t} \sim Cat(\pi_{t})} \!\big[ \ell_{1}(y_{t}, \bar{y}_{t})\nabla_{\!\omega_{A}}\log \pi_{t}(s_{t}) \big]  \Big]  
	+ \alpha\nabla_{\!\omega_{A}} \loss_{2}(\theta, \psi).
	\end{equation}
	\hrulefill \vspace*{1pt} \vspace{-5mm}
\end{figure*}
\subsection{Optimization} \label{sec:optimization}
The optimization problem in \eqref{eq:predictive_cluster_objective} is a non-convex combinatorial problem because it comprises not only minimizing the KL divergence but also finding the optimal cluster assignments and centroids.  
Hence, we propose an optimization procedure that iteratively solves two subproblems: 
i) optimizing the three networks -- the encoder, selector, and predictor -- while fixing the embedding dictionary and ii) optimizing the embedding dictionary while fixing the three networks.
Pseudo-code of \proposed~can be found in the Supplementary Material.

\subsubsection{Optimizing the Three Network}
Finding predictive clusters incorporates the sampling process which is non-differentiable. Thus, to render ``back-propagation'', we utilize the training of actor-critic models \citep{actor_critic:00}. More specifically, we view the combination of the encoder ($f_{\theta}$) and the selector ($h_{\psi}$) as the ``actor'' parameterized by $\omega_{A} = [\theta, \psi]$, and the predictor ($g_{\phi}$) as the ``critic''.
The critic takes as input the the output of the actor (i.e., the cluster assignment) and estimates its value based on the sample-wise predictive clustering loss (i.e., $\ell_{1}(y_{t}, \bar{y}_{t})$) given the chosen cluster. This, in turn, renders the actor to change the distribution of selecting a cluster to minimize such loss.
Thus, it is important for the critic to perform well on the updated output of the actor while it is important for the actor to perform well on the updated loss estimation.
As such, the parameters for the actor and the critic need to be updated iteratively. 


Given the embedding dictionary $\Ec$ fixed (thus, we will omit the dependency on $\Ec$), we train the actor, i.e., the encoder and the selector, by minimizing a combination of the predictive clustering loss $\loss_{1}$ and the entropy of cluster assignments $\loss_{2}$, which is given by $\loss_{A}(\theta, \psi, \phi) = \loss_{1}(\theta, \psi, \phi) + \alpha\loss_{2}(\theta, \psi)$ where $\alpha \geq 0$ is a coefficient chosen to balance between the two losses. 
To derive the gradient of this loss with respect $\omega_{A} = [\theta, \psi]$, we utilize the ideas from actor-critic models \citep{actor_critic:00} in \eqref{eq:loss_1_derivative} which is displayed at the top; please refer to the Supplementary Material for the detailed derivation.
Note that since no sampling process is considered in $\loss_{2}(\theta, \psi)$, we can simply derive $\nabla_{\omega_{A}} \loss_{2}(\theta, \psi)$.

Iteratively with training the actor, we train the critic, i.e., the predictor, by minimizing the predictive clustering loss $\loss_{1}$ as the following:
$\loss_{C}(\phi) = \loss_{1}(\theta, \psi, \phi)$ whose gradient with respect to $\phi$ can be givens as  $\nabla_{\phi} \loss_{C}(\phi) = \nabla_{\phi} \loss_{1}(\theta, \psi, \phi)$.
Note that since the critic is independent of the sampling process, the gradient can be simply back-propagated.

\subsubsection{Optimizing the Cluster Centroids}
Now, once the parameters for the three networks $(\theta, \psi, \phi)$ are fixed (thus, we omit the dependency on $\theta$, $\psi$, and $\phi$),
we updated the embeddings in $\Ec$ by minimizing a combination of the predictive clustering loss $\loss_{1}$ and the embedding separation loss $\loss_{3}$, which is given by 
$\loss_{E}(\Ec) = \loss_{1}(\Ec) + \beta\loss_{3}(\Ec)$ where $\beta \geq 0$ is a coefficient chosen to balance between the two losses.


\subsubsection{Initializing \proposed~via Pre-Training}
Since we transform the combinatorial optimization problem in \eqref{eq:predictive_cluster_objective} into iteratively solving two sub-problems, initialization is crucial to achieve better optimization as a similar concern has been addressed in \citep{Yang:17}.

Therefore, we initialize our model based on the following procedure.
First, we pre-train the encoder and the predictor by minimizing the following loss function based on the predicted label distribution given the latent encodings of input sequences, i.e., $\hat{y}_{t} \triangleq g_{\phi}(\zv_{t})=g_{\phi}(f_{\theta}(\xv_{1:t}))$, as the following:
\begin{equation} \label{eq:loss_prediction_loss}
\loss_{I}(\theta, \phi) = \E_{\xv,y \sim p_{XY}}\Big[- \sum_{t\in\Tc} \ell_{1}(y_{t}, \hat{y}_{t})  \Big].
\end{equation}
Minimizing \eqref{eq:loss_prediction_loss} encourages the latent encoding to be enriched with information for accurately predicting the label distribution.
Then, we perform $K$-means (other clustering method can be also applied) based on the learned representations to initialize the embeddings $\Ec$ and the cluster assignments $\{\{s_{t}^{n}\}_{t=1}^{T^{n}}\}_{n=1}^{N}$. 
Finally, we pre-train the selector $h_{\psi}$ by minimizing the cross entropy treating the initialized cluster assignments as the true clusters.

\section{Related Work}
Temporal clustering, also known as time-series clustering, is a process of unsupervised partitioning of the time-series data into clusters in such a way that homogeneous time-series are
grouped together based on a certain similarity measure. Temporal clustering is challenging because i) the data is often high-dimensional – it consists of sequences not only with high-dimensional features but also with many time points – and ii) defining a proper similarity measure for time-series is not straightforward since it is often highly sensitive to distortions \citep{Ratanamahatana:05}.
To address these challenges, there have been various attempts to find a good representation with reduced dimensionality or to define a proper similarity measure for times-series \citep{Aghabozorgi:15}. 

Recently, \cite{Baytas:17} and \cite{Madiraju:18} proposed temporal clustering methods that utilize low-dimensional representations learned by RNNs.
These works are motivated by the success of applying deep neural networks to find ``clustering friendly'' latent representations for clustering static data \citep{Xie:16,Yang:17}.
In particular, \citet{Baytas:17} utilized a modified LSTM auto-encoder to find the latent representations that are effective to summarize the input time-series and conducted $K$-means on top of the learned representations as an ad-hoc process. 
Similarly, \cite{Madiraju:18} proposed a bidirectional-LSTM auto-encoder that jointly optimizes the reconstruction loss for dimensionality reduction and the clustering objective. 
However, these methods do not associate a target property with clusters and, thus, provide little prognostic value about the underlying disease progression.

Our work is most closely related to SOM-VAE \citep{Fortuin:19}. This method jointly optimizes a static variational auto-encoder (VAE), that finds latent representations of input features, and a self-organizing map (SOM), that allows to map the latent representations into a more interpretable discrete representations, i.e., the embeddings.
However, there are three key differences between our work and SOM-VAE. 
First, SOM-VAE aims at minimizing the reconstruction loss that is specified as the mean squared error between the original input and the reconstructed input based on the corresponding embedding. Thus, similar to the aforementioned methods, SOM-VAE neither associates future outcomes of interest with clusters. In contrast, we focus on minimizing the KL divergence between the outcome distribution given the original input sequence and that given the corresponding embedding to build association between future outcomes of interest and clusters. 
Second, to overcome non-differentiability caused by the sampling process (that is, mapping the latent representation to the embeddings), \cite{Fortuin:19} applies the gradient copying technique proposed by \citep{VQ-VAE:17}, while we utilize the training of actor-critic model \citep{actor_critic:00}.
Finally, while we flexibly model time-series using LSTM, SOM-VAE handles time-series by integrating a Markov model in the latent representations. This can be a strict assumption especially in clinical settings where a patient's medical history is informative for predicting the future clinical outcomes \citep{Ranganath:16}.

\section{Experiments} \label{sec:experiments}
In this section, we provide a set of experiments using two real-world time-series datasets. 
We iteratively update the three networks -- the encoder, selector, and predictor -- and the embedding dictionary as described in Section \ref{sec:optimization}.
For the network architecture, we constructed the encoder utilizing a single-layer LSTM \citep{LSTM:97} with 50 nodes and constructed the selector and predictor utilizing two-layer fully-connected network with 50 nodes in each layer, respectively. 
The parameters $(\theta, \psi, \phi)$ are initialized by Xavier initialization \citep{Xavier:10} and optimized via Adam optimizer \citep{Adam:14} with learning rate of $0.001$ and keep probability $0.7$. 
We chose the balancing coefficients $\alpha, \beta \in \{0.001, 0.01, 0.1, 1.0\}$ utilizing grid search that achieves the minimum validation loss in \eqref{eq:loss_predictive_clustering}; the effect of different loss functions are further investigated in the experiments.
Here, all the results are reported using 5 random 64/16/20 train/validation/test splits. 

\subsection{Real-World Datasets}
We conducted experiments to investigate the performance of \proposed~on two real-world medical datasets; detailed statistics of each dataset can be found in the Supplementary Material.

	\textbf{UK Cystic Fibrosis registry (UKCF)\footnote{\url{https://www.cysticfibrosis.org.uk}}:} 
	This dataset records annual follow-ups for 5,171 adult patients (aged 18 years or older) enrolled in the UK CF registry over the period from 2008 and 2015, with a total of 25,012 hospital visits. 
	Each patient is associated with 89 variables (i.e., 11 static and 78 time-varying features), including information on demographics and genetic mutations, bacterial infections, lung function scores, therapeutic managements, and diagnosis on comorbidities.
	We set the development of different comorbidities in the next year as the label of interest at each time stamp. 
	
	\textbf{Alzheimer's Disease Neuroimaging Initiative (ADNI)\footnote{\url{https://adni.loni.usc.edu}}:} 
	This dataset consists of 1,346 patients in the Alzheimer's disease study with a total of 11,651 hospital visits, which tracks the disease progression via follow-up observations at 6 months interval. 
	Each patient is associated with 21 variables (i.e., 5 static and 16 time-varying features), including information on demographics, biomarkers on brain functions, and cognitive test results.
	We set predictions on the three diagnostic groups -- normal brain functioning, mild cognitive impairment, and Alzheimer's disease -- as the label of interest at each time stamp. 

\subsection{Benchmarks}
We compare \proposed~with clustering methods ranging from conventional approaches based on $K$-means to the state-of-the-art approaches based on deep neural networks. 
All the benchmarks compared in the experiments are tailored to incorporate time-series data as described below:

	\textbf{Dynamic time warping followed by $K$-means}: Dynamic time warping (DTW) is utilized to quantify pairwise distance between two variable-length sequences and, then, $K$-means is applied (\textbf{KM-DTW}). 
	
	\textbf{$K$-means with deep neural networks}: 
	To handle variable-length time-series data, we utilize our encoder and predictor that are trained based on \eqref{eq:loss_prediction_loss} for fixed-length dimensionality reduction.
	Then, we apply $K$-means on the latent encodings $\zv$ (\textbf{KM-E2P ($\Zc$)}) and on the predicted label distributions $\hat{y}$ (\textbf{KM-E2P ($\Yc$)}), respectively.
	
	\textbf{Extensions of DCN} \citep{Yang:17}:  
	Since the DCN is designed for static data, we replace their static auto-encoder with a sequence-to-sequence network to incorporate time-series data (\textbf{DCN-S2S}).\footnote{This extension is a representative of recent deep learning approaches for clustering of both static data \citep{Xie:16,Yang:17} and time-series data \citep{Baytas:17,Madiraju:18} since these methods are built upon the same concept -- that is, applying deep networks for dimensionality reduction to conduct conventional clustering methods, e.g., $K$-means.}
	To associated with the label distribution, we compare a DCN whose static auto-encoder is replaced with our encoder and predictor (\textbf{DCN-E2P}) to focus dimensionality reduction while preserving information for label prediction. 
	
	\textbf{SOM-VAE} \citep{Fortuin:19}: We compare with SOM-VAE -- though, this method aims at visualizing input -- since it naturally clusters time-series data (\textbf{SOM-VAE}).
	In addition, we compare with a variation of SOM-VAE by replacing the decoder with our predictor to find embeddings that capture information for predicting the label (\textbf{SOM-VAE-P}).
	For both cases, we set the dimension of SOM to $K$. 

It is worth highlighting that the label information is provided for training DCN-E2P, KM-E2P, and SOM-VAE-P while the label information is not provided for training KM-DTW, DCN-S2S, and SOM-VAE. 
Please refer to the Supplementary Material for the summary of major components of the tested benchmarks and the implementation details.

\subsection{Performance Metrics}
\textbf{Clustering Performance:} We applied the following three standard metrics for evaluating clustering performances when the ground-truth cluster label is available: \textit{purity score}, \textit{normalized mutual information} (NMI) \citep{metric:NMI}, and \textit{adjusted Rand index} (ARI) \citep{metric:ARI}. 
More specifically, the purity score assesses how homogeneous each cluster is (ranges from 0 to 1 where 1 being a cluster consists of a single class), the NMI is an information theoretic measure of how much information is shared between the clusters and the labels that is adjusted for the number of clusters (ranges from 0 to 1 where 1 being a perfect clustering), and ARI is a corrected-for-chance version of the Rand index which is a measure of the percentage of correct cluster assignments (ranges from -1 to 1 where 1 being a perfect clustering and 0 being a random clustering).

When the ground-truth label is not available, we utilize the average \textit{Silhouette index} (SI) \citep{Silhouette_Index} which measures how similar a member is to its own cluster (homogeneity within a cluster) compared to other clusters (heterogeneity across clusters). Formally, the SI for a subsequence $\xv_{1:t}^{n} \in \Cc^{k}$ can be given as follows: $SI(n) = \frac{b(n) - a(n)}{\max (a(n), b(n))}$ where $a(n) = \frac{1}{|\Cc^{k}|-1} \sum_{m \neq n} \| y_{t}^{n} - y_{t}^{m} \|_{1}$ and $b(n) = \min_{k'\neq k} \frac{1}{|\Cc^{k'}|} \sum_{m \in \Cc^{k'}} \| y_{t}^{n} - y_{t}^{m} \|_{1}$. 
Here, we used the L1-distance between the ground-truth labels of the future outcomes of interest since our goal is to group input subsequences with similar future outcomes.

\textbf{Prediction Performance: } To assess the prediction performance of the identified predictive clusters, we utilized both a\textit{rea under receiver operator characteristic curve} (AUROC) and \textit{area under precision-recall curve} (AUPRC) based on the label predictions of each cluster and the ground-truth binary labels on the future outcomes of interest. Note that the prediction performance is available only for the benchmarks that incorporate the label information during training.

\subsection{Clustering Performance} \label{subsec:clustering_performance}
We start with a simple scenario where the true class (i.e., the ground-truth cluster label) is available and the number of classes is tractable. In particular, we set $C=2^{3}=8$ based on the binary labels for the development of three common comorbidities of cystic fibrosis -- diabetes, ABPA, and intestinal obstruction -- in the next year for the UKCF dataet and $C=3$ based on the mutually exclusive three diagnostic groups for the ADNI dataset. 
We compare \proposed~against the aforementioned benchmarks with respect to the clustering and prediction performance in Table \ref{table:performance_ukcf_and_adni}. 

\begin{table*}[t!]
	\caption{Performance comparison on the UKCF and ADNI datasets.} \label{table:performance_ukcf_and_adni} 
	\begin{center}
		\fontsize{8.0}{9.0}\selectfont
		\begin{tabular}{|c |c| l l l | l l |}
			\hline			
			\textbf{Dataset}&\textbf{Method}& \textbf{Purity} &\textbf{NMI} &\textbf{ARI} &\textbf{AUROC} &\textbf{AUPRC} \\ \cline{1-7}
			\multirow{8}{*}{UKCF}
			&KM-DTW            &0.573$\pm$0.01$^{*}$       &0.010$\pm$0.01$^{*}$     &0.014$\pm$0.01$^{*}$ 
			&N/A &N/A \\
			&KM-E2P ($\Zc$)    &0.719$\pm$0.01$^{*}$       &0.211$\pm$0.01$^{*}$     &0.107$\pm$0.01$^{*}$ 
			&0.726$\pm$0.01$^{*}$  &0.425$\pm$0.02$^{*}$  \\
			&KM-E2P ($\Yc$)    &0.751$\pm$0.01$^{*}$       &0.325$\pm$0.01$^{*}$     &0.440$\pm$0.02$^{*}$ 
			&0.807$\pm$0.00$^{*}$  &0.514$\pm$0.01$^{*}$  \\
			&DCN-S2S           &0.607$\pm$0.06$^{*}$       &0.059$\pm$0.08$^{*}$     &0.063$\pm$0.09$^{*}$ 
			&N/A &N/A \\
			&DCN-E2P           &0.751$\pm$0.02$^{*}$       &0.275$\pm$0.02$^{*}$     &0.184$\pm$0.01$^{*}$ 
			&0.772$\pm$0.03$^{*}$  &0.487$\pm$0.03$^{*}$  \\
			&SOM-VAE           &0.573$\pm$0.01$^{*}$       &0.006$\pm$0.00$^{*}$     &0.006$\pm$0.01$^{*}$ 
			&N/A &N/A \\
			&SOM-VAE-P         &0.638$\pm$0.04$^{*}$       &0.201$\pm$0.05$^{*}$     &0.283$\pm$0.17$^{\dagger}$ 
			&0.754$\pm$0.05$^{*}$  &0.331$\pm$0.07$^{*}$  \\
			&Proposed          &\textbf{0.807$\pm$0.01}    &\textbf{0.463$\pm$0.01}	&\textbf{0.602$\pm$0.01} 
			&\textbf{0.843$\pm$0.01} &\textbf{0.605$\pm$0.01} \\
			\hline 
			\multirow{8}{*}{ADNI}
			&KM-DTW            &0.566$\pm$0.02$^{*}$       &0.019$\pm$0.02$^{*}$     &0.006$\pm$0.02$^{*}$ 
			&N/A &N/A \\
			&KM-E2P ($\Zc$)    &0.736$\pm$0.03$^{\dagger}$ &0.249$\pm$0.02           &0.230$\pm$0.03$^{\dagger}$
			&0.707$\pm$0.01$^{*}$	&0.509$\pm$0.01 \\
			&KM-E2P ($\Yc$)    &0.776$\pm$0.05             &0.264$\pm$0.07           &0.317$\pm$0.11
			&0.756$\pm$0.04  &0.503$\pm$0.04 \\
			&DCN-S2S           &0.567$\pm$0.02$^{*}$       &0.005$\pm$0.00$^{*}$     &0.000$\pm$0.01$^{*}$ 
			&N/A &N/A \\
			&DCN-E2P           &0.749$\pm$0.06             &0.261$\pm$0.05           &0.215$\pm$0.06$^{\dagger}$
			&0.721$\pm$0.03$^{\dagger}$  &0.509$\pm$0.03 \\
			&SOM-VAE           &0.566$\pm$0.02$^{*}$       &0.040$\pm$0.06$^{*}$     &0.011$\pm$0.02$^{*}$
			&N/A &N/A \\
			&SOM-VAE-P         &0.586$\pm$0.06$^{*}$       &0.085$\pm$0.08$^{*}$     &0.038$\pm$0.06$^{*}$
			&0.597$\pm$0.10$^{\dagger}$  &0.376$\pm$0.05$^{*}$ \\
			&Proposed          &\textbf{0.786$\pm$0.03}    &\textbf{0.285$\pm$0.04}  &\textbf{0.330$\pm$0.06}
			&\textbf{0.768$\pm$0.02}	&\textbf{0.515$\pm$0.02}\\	
			\hline
			\multicolumn{6}{l}{$*$ indicates $p\text{-value} < 0.01$,~~~$\dagger$ indicates $p\text{-value} < 0.05$} 
		\end{tabular}
		\vspace{-1mm}
	\end{center}
\end{table*}

As shown in Table \ref{table:performance_ukcf_and_adni}, \proposed~achieved performance gain over all the tested benchmarks in terms of both clustering and prediction performance -- where most of the improvements were statistically significant with $p\text{-value}<0.01$ or $p\text{-value}<0.05$ -- for both datasets. 
Importantly, clustering methods -- i.e., KM-DTW, DCN-S2S, and SOM-VAE -- that do not associate with the future outcomes of interest identified clusters that provide little prognostic value on the future outcomes (note that the true class is derived from the future outcome of interest).
This is clearly shown by the ARI value near 0 which indicates that the identified clusters have no difference with random assignments. 
Therefore, similar sequences with respect to the latent representations tailored for reconstruction or with respect to the shape-based measurement using DTW can have very different outcomes.


\begin{figure*}[t!]
	\centering
	\begin{subfigure}[b]{0.32\linewidth}
		\centering 
		\includegraphics[width=0.95\linewidth, trim= 0.1 0.1 0.1 0.1]{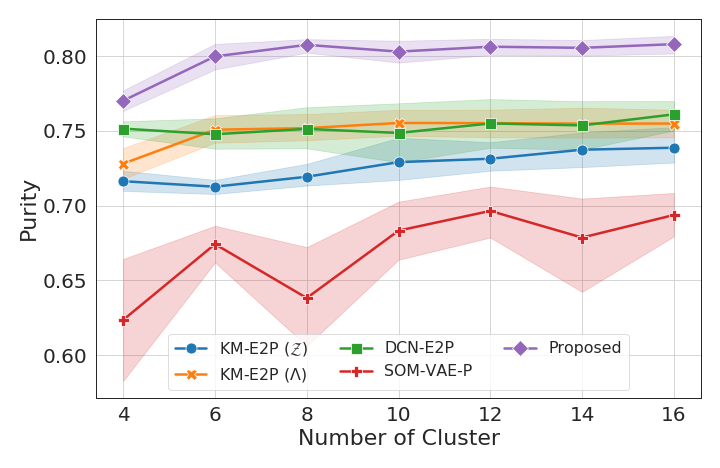} \vspace{-1.0mm}
		\caption{The averaged purity score.} \label{fig:purity} 
	\end{subfigure}	
	\begin{subfigure}[b]{0.32\linewidth}
		\centering 
		\includegraphics[width=0.95\linewidth, trim= 0.1 0.1 0.1 0.1]{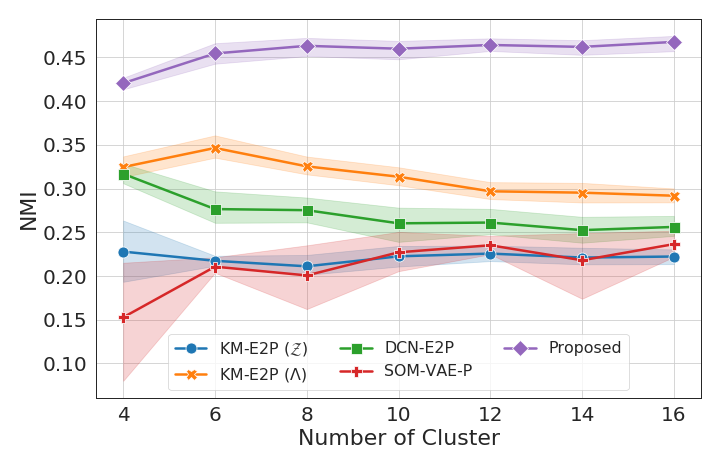} \vspace{-1.0mm}
		\caption{The averaged NMI.} \label{fig:nmi} 
	\end{subfigure}	
	\begin{subfigure}[b]{0.32\linewidth}
		\centering 
		\includegraphics[width=0.95\linewidth, trim= 0.1 0.1 0.1 0.1]{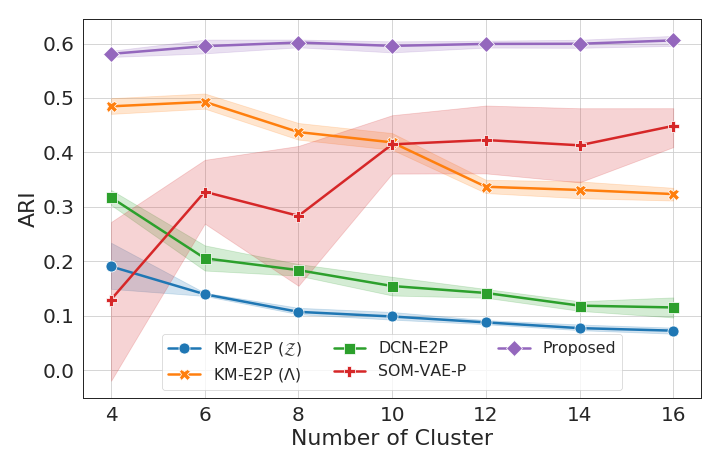} \vspace{-1.0mm}
		\caption{The averaged ARI.} \label{fig:ari} 
	\end{subfigure}
	\caption{The purity score, NMI, and ARI (mean and 95\% confidence interval) for the UKCF dataset ($C=8$) with various $K$.} \vspace{-2mm}
	\label{fig:nmi_ari_ukcf}
\end{figure*}

In Figure \ref{fig:nmi_ari_ukcf}, we further investigate the purity score, NMI, and ARI by varying the number of clusters $K$ from $4$ to $16$ on the UKCF dataset in the same setting with that stated above (i.e., $C=8$). Here, the three methods -- i.e., KM-DTW, DCN-S2S, and SOM-VAE -- are excluded for better visualization.
As we can see in Figure \ref{fig:nmi_ari_ukcf}, our model rarely incur performance loss in both NMI and ARI while the benchmarks (except for SOM-VAE-P) showed significant decrease in the performance as $K$ increased (higher than $C$). 
This is because the number of clusters identified by \proposed~(i.e., the number of activated clusters where we define cluster $k$ is activated if $|\Cc(k)| > 0$) was the same with $C$ most of the times, while the DCN-based methods identified exactly $K$ clusters (due to the $K$-means).
Since the NMI and ARI are adjusted for the number of clusters, a smaller number of identified clusters yields, if everything being equal, a higher performance. 
In contrast, while our model achieved the same purity score for $K\geq8$, the benchmark showed improved performance as $K$ increased since the purity score does not penalize having many clusters.
This is an important property of \proposed~that we do not need to know a priori what the number of cluster is which is a common practical challenge of applying the conventional clustering methods (e.g., $K$-means). 

The performance gain of our model over SOM-VAE-P (and, our analysis is the same for SOM-VAE) comes from two possible sources: i) SOM-VAE-P mainly focuses on visualizing the input with SOM which makes both the encoder and embeddings less flexible -- this is why it performed better with higher $K$ -- and ii) the Markov property can be too strict for time-series data especially in clinical settings where a patient's medical history is informative for predicting the future clinical outcomes \citep{Ranganath:16}. 

\begin{figure*}[t!]
	\centering
	\begin{subfigure}[b]{0.325\linewidth}
		\centering 
		\includegraphics[width=0.92\linewidth, trim= 0.1 0.1 0.1 0.1]{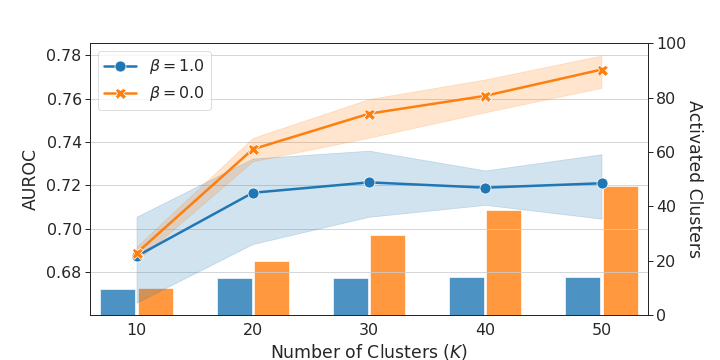} 
		\caption{AUROC} \label{fig:revision_auroc} 
	\end{subfigure}	
	\begin{subfigure}[b]{0.325\linewidth}
		\centering 
		\includegraphics[width=0.92\linewidth, trim= 0.1 0.1 0.1 0.1]{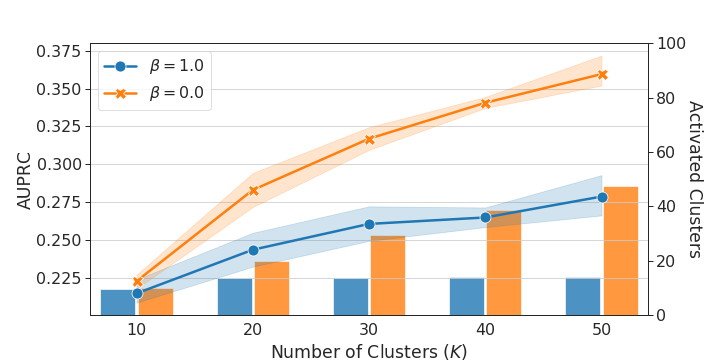} 
		\caption{AUPRC} \label{fig:revision_auprc} 
	\end{subfigure}	
	\begin{subfigure}[b]{0.325\linewidth}
		\centering 
		\includegraphics[width=0.92\linewidth, trim= 0.1 0.1 0.1 0.1]{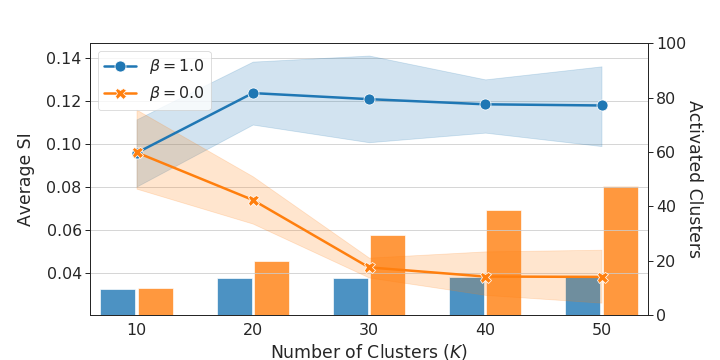} 
		\caption{Average SI} \label{fig:revision_si} 
	\end{subfigure}
	\caption{AUROC, AUPRC, and average SI (mean and 95\% confidence interval) and the number of activated clusters with various $K$.} \vspace{-2mm}
	\label{fig:performance_over_K}
\end{figure*}

\subsection{Multiple Future Outcomes -- a Practical Scenario} \label{subsec:experiments_multiple_outcomes}
In this experiment, we focus on a more practical scenario where the future outcome of interest is high-dimensional and, thus, the number of classes based on all the possible combinations of future outcomes becomes intractable. 
Suppose that we are interested in the development of $M$ comorbidities in the next year whose possible combinations grow exponentially $C=2^{M}$. 
Interpreting such a large number of patient subgroups will be a daunting task which hinders the understanding of underlying disease progression. Since different comorbidities may share common driving factors \citep{comorbidity_ref:2}, we hope our model to identify much smaller underlying (latent) clusters that govern the development of comorbidities.
Here, to incorporate with $M$ comorbidities (i.e., $M$ binary labels), we redefine the output space as $\Yc = \{0,1\}^{M}$ and modify the predictor and loss functions, accordingly. 

\begin{figure}[t!]
	\centering 
	\includegraphics[width=3.3in, trim= 0.1 0.1 0.1 0.1]{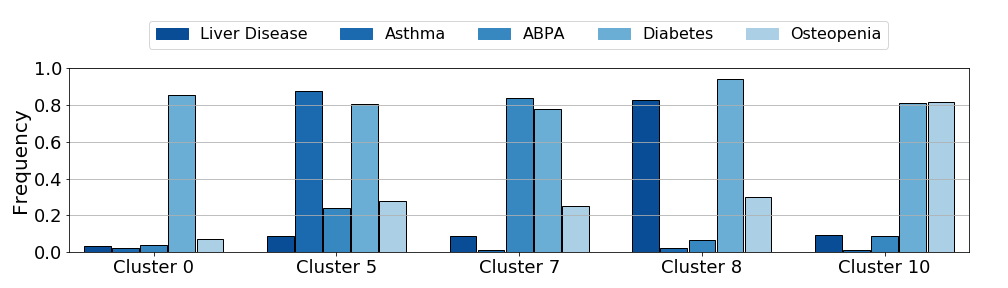}  \vspace{-5mm}
	\caption{Clusters with high-risk of developing diabetes.} \label{fig:ukcf_diabetes_clusters} \vspace{-4mm}
\end{figure}
We identified $12$ clusters of patients based on the next-year development of $22$ different comorbidities in the UKCF dataset and reported $5$ clusters in Figure \ref{fig:ukcf_diabetes_clusters} -- Cluster 0, 5, 7, 8, and 10 -- with the frequency of developing important comorbidities in the next year. Here, we selected the $5$ clusters that have the highest risk of developing diabetes in the next year, and the frequency is calculated in a cluster-specific fashion using the true label. A full list of clusters and comorbidity frequencies can be found in the Supplementary Material.
Although all these clusters displayed high risk of diabetes, the frequency of other co-occurred comorbidities was significantly different across the clusters. 
In particular, around 89\% of the patients in Cluster 5 experienced asthma in the next year while it was less than 3\% of the patients in the other cluster. 
Interestingly, ``leukotriene'' -- a medicine commonly used to manage asthma -- and ``FEV$_{1}$\% predicted'' -- a measure of lung function -- were the two most different input features between patients in Cluster 5 and those in the other clusters.
We observed similar findings in Cluster 7 with ABPA, Cluster 8 with liver disease, and Cluster 10 with osteopenia.
Therefore, by grouping patients who are likely to develop a similar set of comorbidities, our method identified clusters that can be translated into actionable information for clinical decision-making.

\subsection{Trade-Off between Clustering and Prediction}
In predictive clustering, the trade-off between the clustering performance (for better interpretability) -- which quantifies how the data samples are homogeneous within each cluster and heterogeneous across clusters with respect to the future outcomes of interest -- and the prediction performance is a common issue. 
The most important parameter that governs this trade-off is the number of clusters. More specifically, increasing the number of clusters will make the predictive clusters have higher diversity to represent the output distribution and, thus, will increase the prediction performance while decreasing the clustering performance. One extreme example is that there are as many clusters as data samples which will make the identified clusters fully individualized; as a consequence, each cluster will lose interpretability as it no longer groups similar data samples.

To highlight this trade-off, we conduct experiments under the same experimental setup with that of Section \ref{subsec:experiments_multiple_outcomes}.
For the performance measures, we utilized the AUROC and AUPRC to assess the prediction performance, and utilized the average SI to assess the clustering performance. 
To control the number of activated clusters, we set $\beta=0$ and $\beta=1$ (since the embedding separation loss in \eqref{eq:loss_embedding_separation} controls the activation of clusters) and reported the performance by increasing the number of possible clusters $K$, i.e., the dimension of the embedding dictionary.

As can be seen in Figure \ref{fig:performance_over_K}, the prediction performance increased with a increasing number of identified clusters due to the higher diversity to represent the label distribution while making the identified clusters less interpretable. That is, the cohesion and separation among clusters become ambiguous as shown in the low average SI. On the other hand, when we set $\beta=1.0$ (which is selected based on the validation loss in \ref{eq:loss_predictive_clustering}), our method consistently identified a similar number of clusters for $K > 20$, i.e., 13.8 on average, in a data-driven fashion and provided slightly reduced prediction performance with significantly better interpretability, i.e., the average SI $0.120$ on average. 

\section{Conclusion}
In this paper, we introduced \proposed, a deep learning approach for predictive clustering of time-series data. 
We defined novel loss functions to encourage each cluster to have homogeneous future outcomes (e.g., adverse events, the onset of comorbidities, etc.) and designed optimization procedures to avoid trivial solutions in identifying cluster assignments and the centroids.
Throughout the experiments on two real-world datasets, we showed that our model achieves superior clustering performance over state-of-the-art methods and identifies meaningful clusters that can be translated into actionable information for clinical decision-making.


\clearpage

\section*{Acknowledgements}
The authors would like to thank the reviewers for their helpful comments. This work was supported by the National Science Foundation (NSF grants 1524417 and 1722516), the US Office of Naval Research (ONR), and the UK Cystic Fibrosis Trust. We thank Dr. Janet Allen (Director of Strategic Innovation, UK Cystic Fibrosis Trust) for the vision and encouragement. We thank Rebecca Cosgriff and Elaine Gunn for the help with data access, extraction and analysis. We also thank Prof. Andres Floto and Dr. Tomas Daniels, our collaborators, for the very helpful clinical discussions.

\clearpage

%
%
%
%
%

\setcounter{figure}{0}\renewcommand{\thefigure}{S.\arabic{figure}}
\setcounter{table}{0}\renewcommand{\thetable}{S.\arabic{table}}
\setcounter{equation}{0}\renewcommand{\theequation}{S.\arabic{equation}}

	\onecolumn
	\icmltitle{Supplementary Material\\Temporal Phenotyping using Deep Predictive Clustering of Disease Progression}
	
	
	
	
	\begin{icmlauthorlist}
		\icmlauthor{Changhee Lee}{1}
		\icmlauthor{Mihaela van der Schaar}{2,3,1}
	\end{icmlauthorlist}
	
	\icmlaffiliation{1}{University of California, Los Angeles,  USA}
	\icmlaffiliation{2}{University of Cambridge, UK}
	\icmlaffiliation{3}{Alan Turing Institute, UK}
	\icmlcorrespondingauthor{Changhee Lee}{chl8856@ucla.edu}
	
	\icmlkeywords{Machine Learning, ICML}
	
	\vskip 0.3in

	
	
	
	\appendix
	\section{\proposed~for Regression and Binary Classification Tasks} \label{appx:other_labels}
	As the task changes, estimating the label distribution and calculating the KL divergence in (1) of the manuscript must be redefined accordingly:
	For regression task, i.e., $\Yc = \mathbb{R}$, we modify the predictor as $g_{\phi}: \Zc \rightarrow \mathbb{R}$ and replace $\ell_{1}$ by $\ell_{1}(y_{t}, \bar{y}_{t}) = \|y_{t} - \bar{y}_{t} \|_{2}^{2}$. Minimizing $\ell_{1}(y_{t}, \bar{y}_{t})$ is equivalent to minimizing the KL divergence between $p(y_{t}|\xv_{1:t})$ and $p(y_{t}|s_{t})$ when we assume these probability densities follow Gaussian distribution with the same variance.
	For the $M$-dimensional binary classification task, i.e., $\Yc = \{0,1\}^{M}$, we modify the predictor as $g_{\phi}: \Zc \rightarrow [0,1]^{M}$ and replace $\ell_{1}$ by $\ell_{1}(y_{t}, \bar{y}_{t}) = - \sum_{m=1}^{M} y_{t}^{m}\log\bar{y}_{t}^{m} + (1-y_{t}^{m})\log(1-\bar{y}_{t}^{m})$ which is required to minimize the KL divergence. Here, $y_{t}^{m}$ and $\bar{y}_{t}^{m}$ indicate the $m$-th element of $y_{t}$ and $\bar{y}_{t}$, respectively. The basic assumption here is that the distribution of each binary label is independent given the input sequence.

	\section{Detailed Derivation of (5)} 
	To derive the gradient of the predictive clustering loss in (5) of the manuscript with respect $\omega_{A} = [\theta, \psi]$, we utilized the ideas from actor-critic models \citep{actor_critic:00} on $\loss_{A}(\theta, \psi, \phi) = \loss_{1}(\theta, \psi, \phi)$:
	\begin{equation} \label{eq:loss_1_derivative}
	\begin{split}
	\nabla_{\!\omega_{A}} \loss_{A}(\theta, \psi, \phi) 
	&= \E_{\xv,y \sim p_{XY}} \!\left[ \nabla_{\!\omega_{A}} \! \left( \sum_{t=1}^{T} \E_{s_{t} \sim Cat(\pi_{t})}\! \big[ \ell_{1}(y_{t}, \bar{y}_{t}) \big] \right) \right] 
	+ \alpha\nabla_{\!\omega_{A}} \loss_{2}(\theta, \psi) \\
	&= \E_{\xv,y \sim p_{XY}} \!\left[ \sum_{t=1}^{T} \E_{s_{t} \sim Cat(\pi_{t})} \!\big[ \ell_{1}(y_{t}, \bar{y}_{t})\nabla_{\!\omega_{A}}\log \pi_{t}(s_{t}) \big]  \right]  
	+ \alpha\nabla_{\!\omega_{A}} \loss_{2}(\theta, \psi),
	\end{split}
	\end{equation}
	where the second equality comes from the following derivation of the former term:
	\begin{equation} \nonumber
	\begin{split} 
	\E_{\xv,y \sim p_{XY}}\!\left[ \nabla_{\!\omega_{A}} \left( \sum_{t=1}^{T} \E_{s_{t} \sim Cat(\pi_{t})} \big[ \ell_{1}(y_{t}, \bar{y}_{t}) \big] \right) \right]
	&= \E_{\xv,y \sim p_{XY}} \!\left[\nabla_{\!\omega_{A}} \left( \sum_{t=1}^{T}  \sum_{s_{t}\in \Kc}   \pi_{t}(s_{t}) \ell_{1}(y_{t}, \bar{y}_{t}) \right) \right] \\
	&=\E_{\xv,y \sim p_{XY}} \!\left[ \sum_{t=1}^{T} \sum_{s_{t}\in \Kc} \nabla_{\!\omega_{A}} \pi_{t}(s_{t}) \ell_{1}(y_{t}, \bar{y}_{t}) \right] \\
	&=\E_{\xv,y \sim p_{XY}} \!\left[ \sum_{t=1}^{T} \sum_{s_{t}\in \Kc} \frac{\nabla_{\!\omega_{A}} \pi_{t}(s_{t})}{\pi_{t}(s_{t})} \pi_{t}(s_{t}) \ell_{1}(y_{t}, \bar{y}_{t})\right] \\
	&=\E_{\xv,y \sim p_{XY}} \!\left[ \sum_{t=1}^{T} \sum_{s_{t}\in \Kc} \pi_{t}(s_{t}) \ell_{1}(y_{t}, \bar{y}_{t}) \nabla_{\!\omega_{A}}\log \pi_{t}(s_{t}) \right] \\
	&=\E_{\xv,y \sim p_{XY}} \!\left[ \sum_{t=1}^{T} \E_{s_{t} \sim Cat(\pi_{t})} \big[ \ell_{1}(y_{t}, \bar{y}_{t}) \nabla_{\!\omega_{A}}\log \pi_{t}(s_{t}) \big] \right]. \\
	\end{split}
	\end{equation}

	\section{Pseudo-Code of \proposed} \label{appx:pseudo_code}
	As illustrated in Section 3.2, \proposed~is trained in an iterative fashion. We provide the pseudo-code for optimizing our model in Algorithm \ref{alg:pseudo_code} and that for initializing the parameters in Algorithm \ref{alg:pseudo_code_init}. 
	\begin{algorithm*}[t!]
		\caption{Pseudo-code for Optimizing \proposed}
		\label{alg:pseudo_code}
		\begin{algorithmic}
			\footnotesize
			\STATE {\bfseries Input:} Dataset $\data = \{ (\xv_{t}^{n}, y_{t}^{n})_{t=1}^{T^{n}} \}_{n=1}^{N}$, number of clusters $K$, coefficients $(\alpha, \beta)$,\\
			~~~~~~~~~~~~learning rate $(\eta_{A}, \eta_{C}, \eta_{E})$, mini-batch size $n_{mb}$, and update step $M$ 
			\STATE {\bfseries Output:} \proposed~parameters $(\theta, \psi, \phi)$ and the embedding dictionary $\Ec$ 
			\STATE Initialize parameters $(\theta, \psi, \phi)$ and the embedding dictionary $\Ec$ via \texttt{Algorithm \ref{alg:pseudo_code_init}}
			\STATE
			\REPEAT
			\STATE \underline{\textbf{\textit{Optimize the Encoder, Selector, and Predictor}}} 
			\FOR{$m=1, \cdots, M$}
			\STATE Sample a mini-batch of $n_{mb}$ data samples: $\{ (\xv_{t}^{n}, y_{t}^{n})_{t=1}^{T^{n}} \}_{n=1}^{n_{mb}} \sim \data$
			\FOR{$n=1,\cdots, n_{mb}$}
			\STATE Calculate the assignment probability: $~~~\pi_{t}^{n} = [\pi_{t}^{n}(1) \cdots \pi_{t}^{n}(K)] \leftarrow h_{\psi}(f_{\theta}(\xv_{1:t}^{n}))$
			\STATE Draw the cluster assignment: $~~~s^{n}_{t} \sim Cat(\pi_{t}^{n})$
			\STATE Calculate the label distributions: $~~~\bar{y}_{t}^{n} \leftarrow g_{\phi}(\ev(s_{t}^{n}))$ and $\hat{y}_{t}^{n} \leftarrow g_{\phi}(f_{\theta}(\xv_{1:t}^{n}))$
			\ENDFOR
			\STATE Update the encoder $f_{\theta}$ and selector $h_{\psi}$: 
			\begin{equation}
			\begin{split} \nonumber
			\theta &\leftarrow \theta - \eta_{A} \left(  
			\frac{1}{n_{mb}} \sum_{n=1}^{n_{mb}} \sum_{t=1}^{T^{n}} \ell_{1}(y_{t}^{n}, \bar{y}_{t}^{n}) \nabla_{\theta}\log \pi_{t}^{n}(s_{t}^{n}) 
			- \alpha\nabla_{\theta} \sum_{k=1}^{K} \pi_{t}^{n}(k) \log \pi_{t}^{n}(k) 
			\right) \\
			\psi &\leftarrow \psi - \eta_{A} \left(  
			\frac{1}{n_{mb}} \sum_{n=1}^{n_{mb}} \sum_{t=1}^{T^{n}} \ell_{1}(y_{t}^{n}, \bar{y}_{t}^{n}) \nabla_{\psi}\log \pi_{t}^{n}(s_{t}^{n}) 
			- \alpha\nabla_{\psi} \sum_{k=1}^{K} \pi_{t}^{n}(k) \log \pi_{t}^{n}(k) 
			\right)
			\end{split}
			\end{equation}		
			
			\STATE Update the predictor $g_{\phi}$:
			\begin{equation}
			\phi \leftarrow \phi - \eta_{C} \frac{1}{n_{mb}} \sum_{n=1}^{n_{mb}} \sum_{t=1}^{T^{n}} \nabla_{\phi} \ell_{1}(y_{t}^{n}, \bar{y}_{t}^{n}) \nonumber
			\end{equation}
			\ENDFOR
			\STATE   
			\STATE \underline{\textbf{\textit{Optimize the Cluster Centroids}}} \\
			\FOR{$m=1, \cdots, M$}
			\STATE Sample a mini-batch of $n_{mb}$ data samples: $\{ (\xv_{t}^{n}, y_{t}^{n})_{t=1}^{T^{n}} \}_{n=1}^{n_{mb}} \sim \data$
			\FOR{$n=1,\cdots, n_{mb}$}
			\STATE Calculate the assignment probability: $~~~\pi_{t}^{n} = [\pi_{t}^{n}(1) \cdots \pi_{t}^{n}(K)] \leftarrow h_{\psi}(f_{\theta}(\xv_{1:t}^{n}))$
			\STATE Draw the cluster assignment: $~~~s^{n}_{t} \sim Cat(\pi_{t}^{n})$
			\STATE Calculate the label distributions: $~~~\bar{y}_{t}^{n} \leftarrow g_{\phi}(\ev(s_{t}^{n}))$
			\ENDFOR
			\FOR{$k=1, \cdots, K$}
			\STATE Update the embeddings $\ev(k)$:
			\begin{equation} \nonumber
			\ev(k) \leftarrow \ev(k) - \eta_{E} \Bigg( \frac{1}{n_{mb}} \sum_{n=1}^{n_{mb}} \sum_{t=1}^{T^{n}} \nabla_{\ev(k)} \ell_{1}(y_{t}^{n}, \bar{y}_{t}^{n})
			- \gamma 
			\sum_{\substack{k'=1\\k'\neq k}}^{K} \nabla_{\ev(k)} \ell_{1}\big(g_{\phi}(\ev(k)),g_{\phi}(\ev(k'))\big)  
			\Bigg)
			\end{equation}
			\ENDFOR
			\STATE Update the embedding dictionary: $~~~\Ec \leftarrow \{\ev(1), \dots \ev(K)\}$
			\ENDFOR
			\UNTIL convergence
		\end{algorithmic}
	\end{algorithm*}
	\clearpage
	\begin{algorithm*}[t!]
		\caption{Pseudo-code for pre-training \proposed}
		\label{alg:pseudo_code_init}
		\begin{algorithmic}
			\footnotesize
			\STATE {\bfseries Input:} Dataset $\data = \{ (\xv_{t}^{n}, y_{t}^{n})_{t=1}^{T^{n}} \}_{n=1}^{N}$, number of clusters $K$, learning rate $\eta$, mini-batch size $n_{mb}$ 
			\STATE {\bfseries Output:} \proposed~parameters $(\theta, \psi, \phi)$ and the embedding dictionary $\Ec$ 
			\STATE {Initialize parameters $(\theta, \psi, \phi)$ via Xavier Initializer}
			\STATE
			
			\STATE \underline{\textbf{\textit{Pre-train the Encoder and Predictor}}}
			\REPEAT
			\STATE Sample a mini-batch of $n_{mb}$ data samples: $\{ (\xv_{t}^{n}, y_{t}^{n})_{t=1}^{T^{n}} \}_{n=1}^{n_{mb}} \sim \data$
			\FOR{$n=1,\cdots, n_{mb}$}
			\STATE Calculate the label distributions: $~~~\hat{y}_{t}^{n} \leftarrow g_{\phi}(f_{\theta}(\xv_{1:t}^{n}))$
			\ENDFOR
			\begin{equation} \nonumber
			\theta \leftarrow \theta - \eta \frac{1}{n_{mb}} \sum_{n=1}^{n_{mb}} \sum_{t=1}^{T^{n}} \nabla_{\theta} \ell_{1}(y_{t}^{n}, \hat{y}_{t}^{n}) ~~~~~~~~~ \phi \leftarrow \phi - \eta \frac{1}{n_{mb}} \sum_{n=1}^{n_{mb}} \sum_{t=1}^{T^{n}} \nabla_{\phi} \ell_{1}(y_{t}^{n}, \hat{y}_{t}^{n}) 
			\end{equation}
			\UNTIL convergence
			\STATE
			\STATE \underline{\textbf{\textit{Initialize the Cluster Centroids}}}
			\STATE Calculate the embedding dictionary $\Ec$ and initial cluster assignments $c_{t}^{n}$
			\begin{equation} \nonumber
			\Ec, \{\{ c_{t}^{n} \}_{t=1}^{T^{n}} \}_{n=1}^{N} \leftarrow \texttt{K-means}(\{\{ \zv_{t}^{n} \}_{t=1}^{T^{n}} \}_{n=1}^{N}, K)
			\end{equation}
			\STATE \underline{\textbf{\textit{Pre-train the Selector}}}
			\REPEAT
			\STATE Sample a mini-batch of $n_{mb}$ data samples: $\{ (\xv_{t}^{n}, y_{t}^{n})_{t=1}^{T^{n}} \}_{n=1}^{n_{mb}} \sim \data$
			\FOR{$n=1,\cdots, n_{mb}$}
			\STATE Calculate the assignment probability: $~~~\pi_{t}^{n} = [\pi_{t}^{n}(1) \cdots \pi_{t}^{n}(K)] \leftarrow h_{\psi}(f_{\theta}(\xv_{1:t}^{n}))$
			\ENDFOR
			\STATE Update the selector $h_{\psi}$: 
			\begin{equation} \nonumber
			\psi \leftarrow \psi + \eta   
			\frac{1}{n_{mb}} \sum_{n=1}^{n_{mb}} \sum_{t=1}^{T^{n}} \sum_{k=1}^{K} c_{t}^{n}(k) \log \pi_{t}^{n}(k) 
			\end{equation}		
			\UNTIL convergence
		\end{algorithmic}
	\end{algorithm*}

	\begin{table*}[t!]
		\caption{Summary and description of the UKCF dataset.} \label{table:statistics_ukcf}
		\scriptsize
		\begin{center}
			\begin{tabular}{l l l l l l l l l }
				\multicolumn{2}{l}{\textbf{STATIC COVARIATES}} &\textbf{Type}&\textbf{Mean}&\textbf{} & &\textbf{Type}&\textbf{Mean}&\textbf{}\\ \toprule[0.4mm]
				\textbf{Demographic}
				&Gender&Bin.&0.55&  & & & &  \\ \hline							
				\textbf{Genetic}
				&Class I Mutation	&Bin.&0.05& 			&Class VI Mutation	&Bin.	&0.86&	 \\	
				&Class II Mutation	&Bin.&0.87& 			&DF508 Mutation		&Bin.	&0.87&	 \\	
				&Class III Mutation	&Bin.&0.89& 			&G551D Mutation		&Bin.	&0.06&	 \\	
				&Class IV Mutation	&Bin.&0.05& 			&Homozygous			&Bin.	&0.58&	 \\	
				&Class V Mutation	&Bin.&0.04& 			&Heterozygous		&Bin	&0.42&	 \\ \hline
				& & & & & & & &  \\ 
				& & & & & & & &  \\ 
				\multicolumn{2}{l}{\textbf{TIME-VARYING COVARIATES}} &\textbf{Type}&\textbf{\makecell{Mean}}&\textbf{Min / Max} & &\textbf{Type}&\textbf{\makecell{Mean}}&\textbf{Min / Max} \\ 
				\toprule[0.4mm]
				\textbf{Demographic}	
				&Age				&Cont.	&30.4	&18.0 / 86.0	&Height		&Cont.	&168.0	&129.0 / 198.6 \\
				&Weight				&Cont.	&64.1	&24.0 / 173.3	&BMI		&Cont.	&22.6	&10.9 / 30.0 \\
				&Smoking Status		&Bin.	&0.1	& & & & &  \\ \hline	
				\textbf{Lung Func. Scores}
				&FEV$_{1}$				&Cont.	&2.3	&0.2 / 6.3	&Best FEV$_{1}$				&Cont.	&2.5	&0.3 / 8.0 \\
				&FEV$_{1}$\% Pred.  &Cont.	&65.1	&9.0 / 197.6 	&Best FEV$_{1}$\% Pred.	&Cont.	&71.2	&7.5 / 164.3 \\ \hline
				\textbf{Hospitalization}	
				&IV ABX Days Hosp.	&Cont.	&12.3	&0 / 431&Non-IV Hosp. Adm.	&Cont.	&1.2	&0 / 203 \\
				&IV ABX Days Home	&Cont.	&11.9	&0 / 441& & & &\\ \hline
				\textbf{Lung Infections} 
				&B. Cepacia	    &Bin.	&0.05 &    &P. Aeruginosa	&Bin.	&0.59 &  \\	
				&H. Influenza	&Bin.	&0.05 & 	&K. Pneumoniae	&Bin.	&0.00 &  \\
				&E. Coli	    &Bin.	&0.01 & 	&ALCA			&Bin.	&0.03 &  \\	
				&Aspergillus	&Bin.	&0.14 & 	&NTM			&Bin.	&0.03 &  \\	
				&Gram-Negative	&Bin.	&0.01 & 	&Xanthomonas	&Bin.	&0.05 &  \\		
				&S. Aureus	    &Bin.	&0.30 & 	& & & &  \\ \hline
				\textbf{Comorbidities}	
				&Liver Disease	&Bin.	&0.16 &   &Depression				&Bin.	&0.07 &  \\		
				&Asthma			&Bin.	&0.15 &   &Hemoptysis				&Bin.	&0.01 &  \\		
				&ABPA			&Bin.	&0.12 &   &Pancreatitus			&Bin.	&0.01 &  \\		
				&Hypertension	&Bin.	&0.04 &   &Hearing Loss			&Bin.	&0.03 &  \\		
				&Diabetes		&Bin.	&0.28 &   &Gall bladder			&Bin.	&0.01 &  \\		
				&Arthropathy	&Bin.	&0.09 &   &Colonic structure		&Bin.	&0.00 &  \\		
				&Bone fracture	&Bin.	&0.01 &   &Intest. Obstruction	    &Bin.	&0.08 &  \\		
				&Osteoporosis	&Bin.	&0.09 &   &GI bleed -- no var.	    &Bin.	&0.00 &  \\		
				&Osteopenia		&Bin.	&0.21 &   &GI bleed -- var.        &Bin.	&0.00 &  \\		
				&Cancer			&Bin.	&0.00 &   &Liver Enzymes			&Bin.	&0.16 &  \\		
				&Cirrhosis		&Bin.	&0.03 &   &Kidney Stones	        &Bin.	&0.02 &  \\ \hline
				\textbf{Treatments}	
				&Dornase Alpha			&Bin.	&0.56 & 	&Inhaled B. BAAC	&Bin.	&0.03 & 	\\	
				&Anti-fungals			&Bin.	&0.07 & 	&Inhaled B. LAAC	&Bin.	&0.08 & 	\\
				&HyperSaline			&Bin.	&0.23 & 	&Inhaled B. SAAC	&Bin.	&0.05 & 	\\
				&HypertonicSaline		&Bin.	&0.01 & 	&Inhaled B. LABA	&Bin.	&0.11 & 	\\
				&Tobi Solution			&Bin.	&0.20 & 	&Inhaled B. Dilators&Bin.	&0.57 & 	\\
				&Cortico Combo			&Bin.	&0.41 & 	&Cortico Inhaled	&Bin.	&0.15 & 	\\
				&Non-IV Ventilation		&Bin.	&0.05 & 	&Oral B. Theoph.	&Bin.	&0.04 & 	\\
				&Acetylcysteine			&Bin.	&0.02 & 	&Oral B. BA			&Bin.	&0.03 & 	\\
				&Aminoglycoside			&Bin.	&0.03 & 	&Oral Hypo. Agents	&Bin.	&0.01 & 	\\
				&iBuprofen				&Bin.	&0.00 & 	&Chronic Oral ABX   &Bin.	&0.53 & 	\\	
				&Drug Dornase			&Bin.	&0.02 & 	&Cortico Oral			&Bin.	&0.14 & 	\\
				&HDI Buprofen			&Bin.	&0.00 & 	&Oxygen Therapy			&Bin.	&0.11 & 	\\
				&Tobramycin				&Bin.	&0.03 & 	&O$_{2}$ Exacerbation	&Bin.	&0.03 & 	\\
				&Leukotriene			&Bin.	&0.07 & 	&O$_{2}$ Nocturnal		&Bin.	&0.03 & 	\\
				&Colistin				&Bin.	&0.03 & 	&O$_{2}$ Continuous		&Bin.	&0.03 & 	\\
				&Diabetes Insulin		&Bin.	&0.01 & 	&O$_{2}$ Pro re nata	&Bin.	&0.01 & 	\\
				&Macrolida ABX	        &Bin.	&0.02 & 	& & & &  \\ \hline
				\multicolumn{3}{l}{ABX: antibiotics}
			\end{tabular}
		\end{center}
	\end{table*}
	\begin{table*}[t!]
		\caption{Summary and description of the ADNI dataset.} \label{table:statistics_adni}
		\scriptsize
		\begin{center}
			\begin{tabular}{l l l l l l l l l }
				\multicolumn{2}{l}{\textbf{STATIC COVARIATES}} &\textbf{Type}&\textbf{\makecell{Mean }}&\textbf{\makecell{Min/Max (Mode)}}  & &\textbf{Type}&\textbf{\makecell{Mean}}&\textbf{\makecell{Min/Max (Mode)}} \\ \toprule[0.4mm]
				\textbf{Demographic}
				&Race   	    &Cat.&0.93& White 	&Ethnicity		&Cat.	&0.97& No Hisp/Latino \\	
				&Education  	&Cat.&0.23& C16 	&Marital Status	&Cat.	&0.75& Married	 \\	\hline
				\textbf{Genetic}
				&APOE$_{4}$	 	&Cont.	&0.44	&0/2   & & & &  \\ \hline	
				& & & & & & & &  \\ 
				& & & & & & & &  \\ 
				\multicolumn{2}{l}{\textbf{TIME-VARYING COVARIATES}}
				&\textbf{Type}&\textbf{Mean}&\textbf{Min / Max} & &\textbf{Type}&\textbf{Mean}&\textbf{Min / Max} \\ 
				\toprule[0.4mm]
				\textbf{Demographic}	
				&Age			&Cont.	&73.6	&55/92 & & & &  \\ \hline	
				\textbf{Biomarker} 
				&Entorhinal	    &Cont.	&3.6E+3 &1.0E+3 / 6.7E+3    &Mid Temp 	 &Cont.	&2.0E+4 &8.9E+3 / 3.2E+4  \\	
				&Fusiform	    &Cont.	&1.8E+5 &9.0E+4 / 2.9E+5 	&Ventricles	 &Cont.	&4.1E+4 &5.7E+3 / 1.6E+5  \\
				&Hippocampus	&Cont.	&6.9E+3 &2.8E+3 / 1.1E+4 	&Whole Brain &Cont.	&1.0E+6 &6.5E+5 / 1.5E+6  \\	
				&Intracranial	&Cont.	&1.5E+6 &2.9E+2 / 2.1E+6 	& & & &  \\ \hline
				\textbf{Cognitive}	
				&ADAS-11     	  &Cont.	&8.58 &0/70   &ADAS-13			 &Cont.	&13.61 &0/85  \\		
				&CRD Sum of Boxes &Cont.	&1.21 &0/17   &Mini Mental State &Cont.	&27.84 &2/30  \\		
				&RAVLT Forgetting &Cont.	&4.19 &-12/15 &RAVLT Immediate   &Cont.	&38.25 &0/75  \\		
				&RAVLT Learning	  &Cont.	&4.65 &-5/14  &RAVLT Percent     &Cont.	&51.70 &-500/100  \\ \hline
			\end{tabular}
		\end{center}
	\end{table*}
	\section{Details of the Datasets} \label{appx:datasets}
	\subsection{UKCF Dataset}
	UK Cystic Fibrosis registry (UKCF)\footnote{\url{https://www.cysticfibrosis.org.uk/the-work-we-do/uk-cf-registry}} records annual follow-ups for 5,171 adult patients (aged 18 years or older) over the period from 2008 and 2015, with a total of 25,012 hospital visits. 
	Each patient is associated with 89 variables (i.e., 11 static and 78 time-varying features), including information on demographics and genetic mutations, bacterial infections, lung function scores, therapeutic managements, and diagnosis on comorbidities.
	The detailed statistics are given in Table \ref{table:statistics_ukcf}.
	
	\subsection{ADNI Dataset}
	Alzheimer's Disease Neuroimaging Initiative (ADNI)\footnote{\url{https://adni.loni.usc.edu}} study consists of 1,346 patients with a total of 11,651 hospital visits, which tracks the disease progression via follow-up observations at 6 months interval. 
	Each patient is associated with 21 variables (i.e., 5 static and 16 time-varying features), including information on demographics, biomarkers on brain functions, and cognitive test results. 
	The three diagnostic groups were normal brain functioning ($0.55$), mild cognitive impairment ($0.43$), and Alzheimer's disease ($0.02$).
	The detailed statistics are given in Table \ref{table:statistics_adni}.

	\begin{figure*}[t!]
		\centering
		\begin{subfigure}[b]{0.24\linewidth}
			\centering 
			\includegraphics[width=0.6\linewidth, trim= 0.1 0.1 0.1 0.1]{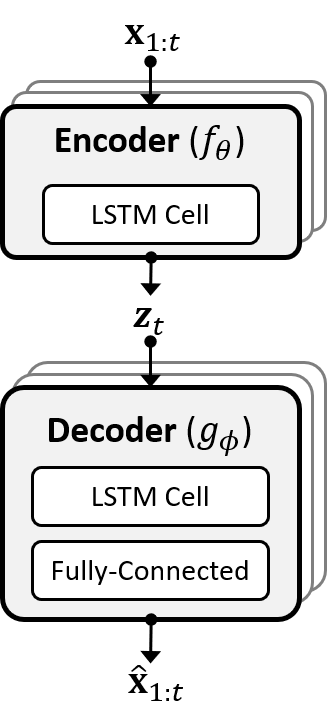}
			\caption{DCN-S2S} \label{fig:network_architecture_dcn_s2s} 
		\end{subfigure}
		\begin{subfigure}[b]{0.24\linewidth}
			\centering 
			\includegraphics[width=0.6\linewidth, trim= 0.1 0.1 0.1 0.1]{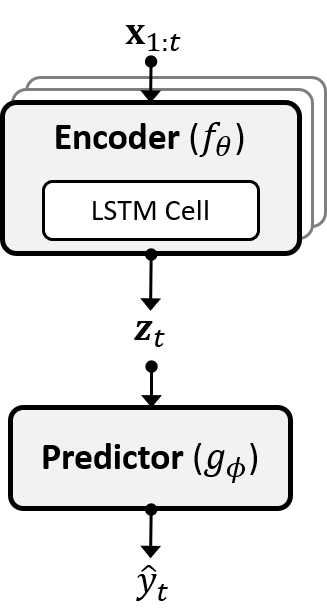}
			\caption{DCN-E2P, KM-E2P} \label{fig:network_architecture_dcn_e2p} 
		\end{subfigure}
		\begin{subfigure}[b]{0.21\linewidth}
			\centering 
			\includegraphics[width=0.6\linewidth, trim= 0.1 0.1 0.1 0.1]{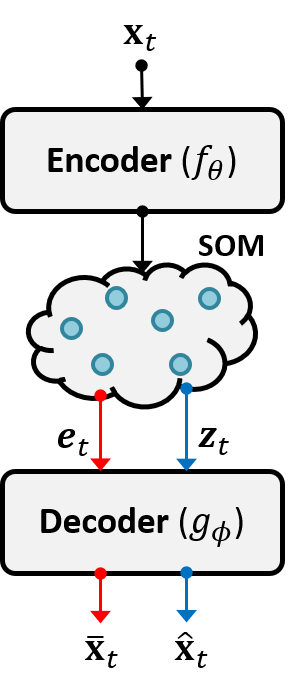}
			\caption{SOM-VAE} \label{fig:network_architecture_som_vae} 
		\end{subfigure}
		\begin{subfigure}[b]{0.21\linewidth}
			\centering 
			\includegraphics[width=0.6\linewidth, trim= 0.1 0.1 0.1 0.1]{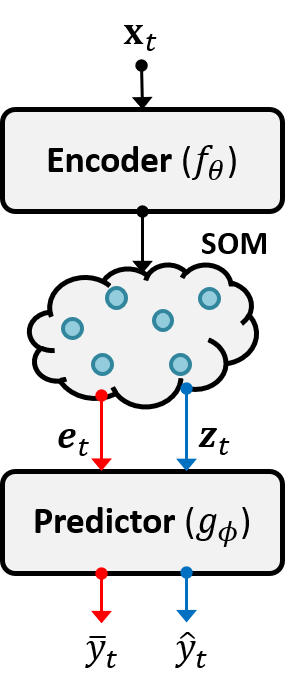}
			\caption{SOM-VAE-P} \label{fig:network_architecture_som_vae_p} 
		\end{subfigure}
		\caption{The block diagrams of the tested benchmarks.}
		\label{fig:network_achitecture_benchmarks}
	\end{figure*}
	\begin{table*}[t!]
		\caption{Comparison table of benchmarks.}\label{table:benchmark_comparison}
		\begin{center}
			\fontsize{8.4}{9.5}\selectfont
			\begin{tabular}{| c | c c c c c|}
				\hline
				\textbf{Methods} &\textbf{\makecell{Handling\\Time-Series}}& \textbf{\makecell{Clustering\\Method}}& \textbf{\makecell{Similarity\\Measure}}& \textbf{\makecell{Label\\Provided}}& \textbf{\makecell{Label\\Associated}} \\ \hline
				KM-DTW         &DTW	         &$K$-means         &DTW	             &N &N             \\
				KM-E2P ($\Zc$) &RNN	         &$K$-means	        &Euclidean in $\Zc$	 &Y &Y (indirect)  \\
				KM-E2P ($\Yc$) &RNN	         &$K$-means	        &Euclidean in $\Yc$	 &Y &Y (direct)    \\
				DCN-S2S	       &RNN	         &$K$-means	        &Euclidean in $\Zc$	 &N &N             \\
				DCN-E2P	       &RNN	         &$K$-means	        &Euclidean in $\Zc$  &Y &Y (indirect)  \\
				SOM-VAE	       &Markov model &embedding mapping &reconstruction loss &N &N	           \\
				SOM-VAE-P	   &Markov model &embedding mapping	&prediction loss     &Y &Y (direct)    \\
				Proposed	   &RNN          &embedding mapping	&KL divergence       &Y &Y (direct)    \\
				\hline
			\end{tabular}
		\end{center}
	\end{table*}
	
	\section{Details of the Benchmarks} \label{appx:benchmarks}
	We compared \proposed~in the experiments with clustering methods ranging from conventional approaches based on $K$-means to the state-of-the-art approaches based on deep neural networks. 
	The details of how we implemented the benchmarks are described as the following: \vspace{-2mm}
	\begin{itemize}[leftmargin=1.5em]
		\item \textbf{Dynamic time warping followed by $K$-means}\footnote{\url{https://github.com/rtavenar/tslearn}}: Dynamic time warping (DTW) is utilized to quantify pairwise distance between two variable-length sequences and, then, $K$-means is applied (denoted as \textbf{KM-DTW}). 
		
		\item \textbf{$K$-means with deep neural networks}: 
		To handle variable-length time-series data, we utilized an encoder-predictor network as depicted in Figure \ref{fig:network_architecture_dcn_e2p} and trained the network based on (6) for dimensionality reduction; this is to provide fixed-length and low-dimensional representations for time-series.
		Then, we applied $K$-means on the latent encodings $\zv$ (denoted as \textbf{KM-E2P ($\Zc$)}) and on the predicted label distributions $\hat{y}$ (denoted as \textbf{KM-E2P ($\Yc$)}), respectively. We implemented the encoder and predictor of KM-E2P with the same network architectures with those of our model: the encoder is a single-layer LSTM with 50 nodes and the decoder is a two-layered fully-connected network with 50 nodes in each layer.	
		
		\item \textbf{Extensions of DCN\footnote{\url{https://github.com/boyangumn/DCN}}} \citep{Yang:17}: 
		Since the DCN is designed for static data, we utilized a sequence-to-sequence model in Figure \ref{fig:network_architecture_dcn_s2s} for the encoder-decoder network as an extension to incorporate time-series data (denoted as \textbf{DCN-S2S}) and trained the network based on the reconstruction loss (using the reconstructed input sequence $\hat{\xv}_{1:t}$). For implementing DCN-S2S, we used a single-layer LSTM with 50 nodes for both the encoder and the decoder. And, we augmented a fully-connected layer with 50 nodes is used to reconstruct the original input sequence from the latent representation of the decoder.	
		
		In addition, since predictive clustering is associated with the label distribution, we compared a DCN whose encoder-decoder structure is replaced with our encoder-predictor network in Figure \ref{fig:network_architecture_dcn_e2p} (denoted as \textbf{DCN-E2P}) to focus the dimensionality reduction -- and, thus, finding latent encodings where clustering is performed -- on the information for predicting the label distribution. We implemented the encoder and predictor of DCN-E2P with the same network architectures with those of our model as described in Section 5.
		
		\item \textbf{SOM-VAE}\footnote{\url{https://github.com/ratschlab/SOM-VAE}} \citep{Fortuin:19}: We compare with SOM-VAE -- though, this method is oriented towards visualization of input data via SOM -- since it naturally clusters time-series data assuming Markov property (denoted as \textbf{SOM-VAE}). We replace the original CNN architecture of the encoder and the decoder with three-layered fully-connected network with 50 nodes in each layer, respectively. The network architecture is depicted in Figure \ref{fig:network_architecture_som_vae} where $\hat{\xv}_{t}$ and $\bar{\xv}_{t}$ indicate the reconstructed inputs based on the encoding $\zv_{t}$ and the embedding $\ev_{t}$ at time $t$, respectively.
		
		In addition, we compare with a variation of SOM-VAE by replacing the decoder with the predictor to encourage the latent encoding to capture information for predicting the label distribution (denoted as \textbf{SOM-VAE-P}). For the implementation, we replaced the decoder of SOM-VAE with our predictor which is a two-layered fully-connected layer with 50 nodes in each layer to predict the label distribution as illustrated in Figure \ref{fig:network_architecture_som_vae_p}. Here, $\hat{y}_{t}$ and $\bar{y}_{t}$ indicate the predicted labels based on the encoding $\zv_{t}$ and the embedding $\ev_{t}$ at time $t$, respectively.
		
		For both cases, we used the default values for balancing coefficients of SOM-VAE and the dimension of SOM to be equal to $K$. 
	\end{itemize} \vspace{-2mm}
	We compared and summarized major components of the benchmarks in Table \ref{table:benchmark_comparison}.

	\section{Additional Experiments} \label{appx:additional_results}
	\subsection{Contributions of the Auxiliary Loss Functions}
	As described in Section 3.1, we introduced two auxiliary loss functions -- the sample-wise entropy of cluster assignment (3) and the embedding separation loss (4) -- to avoid trivial solution that may arise in identifying the predictive clusters. 
	To analyze the contribution of each auxiliary loss function, we report the average number of activated clusters, clustering performance, and prediction performance on the UKCF dataset with $3$ comorbidities as described in Section 5.4. 
	Throughout the experiment, we set $K=16$ -- which is larger than $C$ -- to find the contribution of these loss functions to the number of activated clusters. 
	\begin{table*}[h!]
		\caption{Performance comparison with varying the balancing coefficients $\alpha, \beta$ for the UKCF dataset.} \label{table:sensitivity_to_coeffcients}
		\begin{center}
			\small
			\begin{tabular}{| c c | c  c c c | c c |}
				\hline			
				\multicolumn{2}{|c|}{\underline{\textbf{~Coefficients~}}} & \multicolumn{4}{c|}{\underline{\textbf{~~~~~~~~~~~~Clustering Performance~~~~~~~~~~~~}}} & \multicolumn{2}{c|}{\underline{\textbf{Prognostic Value}}}  \\ 
				$\alpha$&$\beta$ & Activated No. &Purity &NMI &ARI &AUROC &AUPRC \\ \hline
				0.0 &0.0 &16  &0.573$\pm$0.01 &0.006$\pm$0.00 &0.000$\pm$0.00 &0.500$\pm$0.00 &0.169$\pm$0.00 \\
				0.0 &1.0 &16  &0.573$\pm$0.01 &0.006$\pm$0.00 &0.000$\pm$0.00 &0.500$\pm$0.00 &0.169$\pm$0.00 \\
				3.0 &0.0 &8.4   &0.795$\pm$0.01 &0.431$\pm$0.01 &0.569$\pm$0.01 &0.840$\pm$0.01 &0.583$\pm$0.02 \\
				3.0 &1.0 &8 &\textbf{0.808$\pm$0.01} &\textbf{0.468$\pm$0.01} &\textbf{0.606$\pm$0.01} &\textbf{0.852$\pm$0.00} &\textbf{0.608$\pm$0.01} \\
				\hline
			\end{tabular}
		\end{center}
	\end{table*}
	
	As we can see in Table \ref{table:sensitivity_to_coeffcients}, both auxiliary loss functions make important contributions in improving the quality of predictive clustering. 
	More specifically, the sample-wise entropy encourages the selector to choose one dominant cluster. Thus, as we can see results with $\alpha=0$, without the sample-wise entropy, our selector assigns an equal probability to all $16$ clusters which results in a trivial solution. 
	We observed that, by augmenting the embedding separation loss (4), \proposed~identifies a smaller number of clusters owing to the well-separated embeddings.

	\subsection{Additional Results on Targeting Multiple Future Outcomes}
	Throughout the experiment in Section 5.5, we identified $12$ subgroups of patients that are associated with the next-year development of $22$ different comorbidities in the UKCF dataset. 
	In Table \ref{table:cluster_statistics_all}, we reported $12$ identified clusters and the full list of comorbidities developed in the next year since the latest observation and the corresponding frequency which is calculated in a cluster-specific fashion based on the true label.
	
	As we can see in the table, the identified clusters displayed very different label distributions; that is, the combination of comorbidities as well as their frequency were very different across the clusters. For example, patients in Cluster 1 experienced low-risk of developing any comorbidities in the next year while 85\% of patients in Cluster 0 experienced diabetes in the next year.
	
	\begin{table*}[t!]
		\caption{The comorbidities developed in the next year for the $12$ identified clusters. The values in parentheses indicate the corresponding frequency.}\label{table:cluster_statistics_all}
		\begin{center}
			\scriptsize
			\begin{tabular}{| c | l l l l|}
				\hline
				\textbf{Clusters} & \multicolumn{4}{c|}{\textbf{Comorbidities and Frequencies}} \\ \hline
				\multirow{6}{*}{\makecell{Cluster\\0}}&	Diabetes (0.85)& Liver Enzymes (0.21)& Arthropathy (0.14)& Depression (0.10)\\
				& Hypertens (0.08)& Osteopenia (0.07)& Intest. Obstruction (0.07)& Cirrhosis (0.04)\\
				&	ABPA (0.04)& Liver Disease (0.04)& Osteoporosis (0.03)& Hearing Loss (0.03) \\
				&	Asthma (0.02)& Kidney Stones (0.01)& Bone fracture (0.01)&Hemoptysis (0.01) \\
				&	Pancreatitis (0.01)& Cancer (0.00)& Gall bladder (0.00)& Colonic stricture (0.00)\\
				& GI bleed -- no var. (0.00)& GI bleed -- var. (0.00)&  & \\ \hline
				\multirow{6}{*}{\makecell{Cluster\\1}}& Liver Enzymes (0.09)& Arthropathy (0.08)& Depression (0.07)& Intest. Obstruction (0.06) \\
				&	Diabetes (0.06)& Osteopenia (0.05)& ABPA (0.04)& Asthma	(0.03) \\
				&	Liver Disease (0.03)& Hearing Loss (0.03)& Osteoporosis	(0.02)& Pancreatitis (0.02)\\
				& Kidney Stones (0.02)& Hypertension (0.01)& Cirrhosis (0.01)& Gall bladder (0.01)\\
				&	Cancer (0.01)& Hemoptysis (0.00)& Bone fracture (0.00)& Colonic stricture (0.00) \\
				&	GI bleed -- no var. (0.00)& GI bleed -- var. (0.00)& & \\ \hline
				\multirow{6}{*}{\makecell{Cluster\\2}}& ABPA (0.77)& Osteopenia	(0.21)& Intest. Obstruction	(0.11)& Hearing Loss (0.10) \\
				& Liver Enzymes (0.07)& Diabetes (0.06)& Depression (0.05)& Pancreatitis (0.05) \\
				&	Liver Disease (0.04)& Arthropathy (0.04)& Asthma (0.03)& Bone fracture (0.02) \\
				&	Osteoporosis (0.02)& Hypertension (0.01)& Cancer (0.01)& Cirrhosis (0.01)\\
				&	Kidney Stones (0.01)& Gall bladder (0.01)& Hemoptysis (0.00)& Colonic stricture (0.00)\\
				& GI bleed -- no var. (0.00)&	GI bleed -- var. (0.00)& & \\ \hline
				\multirow{6}{*}{\makecell{Cluster\\3}}& Asthma (0.89)& Liver Disease (0.87)& Diabetes (0.29)& Osteopenia (0.28) \\
				& Liver Enzymes (0.24)& ABPA (0.15)& Osteoporosis (0.11)& Hearing Loss (0.06) \\
				& Arthropathy (0.05)& Intest. Obstruction (0.05)& Depression (0.04)& Hypertension (0.03)\\
				& Cirrhosis (0.02)& Kidney Stones (0.02)&	Pancreatitis (0.02)& Gall bladder (0.02) \\
				& Cancer (0.01)& Bone fracture (0.00)& Hemoptysis (0.00)& Colonic stricture (0.00) \\
				& GI bleed -- no var. (0.00)& GI bleed -- var. (0.00)& & \\ \hline
				\multirow{6}{*}{\makecell{Cluster\\4}}& Osteoporosis (0.76)& Diabetes (0.43)& Arthropathy (0.20)& Liver Enzymes	(0.18) \\
				& Osteopenia (0.15)& Depression (0.13)& Intest. Obstruction (0.11)& ABPA (0.11) \\
				&	Hearing Loss (0.09)& Liver Disease (0.08)& Hypertension	(0.07)& Cirrhosis (0.07) \\
				&	Kidney Stones (0.03)& Asthma (0.02)& Hemoptysis	(0.02)& Bone fracture (0.02)\\
				&	Gall bladder (0.01)& Pancreatitis (0.01)& Cancer (0.00)& Colonic stricture (0.00) \\
				&	GI bleed -- no var.	(0.00)&	GI bleed -- var. (0.00)& & \\ \hline
				\multirow{6}{*}{\makecell{Cluster\\5}}& Asthma (0.88)& Diabetes (0.81)& Osteopenia (0.28)& ABPA (0.24)\\
				& Liver Enzymes (0.22)& Depression (0.15)& Osteoporosis (0.14)& Intest. Obstruction (0.12) \\
				& Hypertension (0.10)& Cirrhosis (0.10)& Liver Disease (0.09)& Arthropathy (0.08) \\
				& Bone fracture (0.01)& Hemoptysis (0.01)& Pancreatitis (0.01)& Hearing Loss (0.01) \\
				&	Cancer (0.01)& Kidney Stones (0.01)& GI bleed -- var. (0.01)& Gall bladder (0.00) \\
				& Colonic stricture (0.00)& GI bleed -- no var. (0.00)& & \\ \hline
				\multirow{6}{*}{\makecell{Cluster\\6}}& Liver Disease (0.85)& Liver Enzymes (0.37)& Osteopenia (0.27)& ABPA (0.09)\\
				& Arthropathy (0.07)& Diabetes (0.06)& Intest. Obstruction (0.06)& Osteoporosis (0.05) \\
				& Depression (0.03)& Asthma (0.03)& Hearing Loss (0.03)& Cirrhosis (0.02) \\
				& Hemoptysis (0.02)& Hypertension (0.01)& Kidney Stones (0.01)& Pancreatitis (0.00) \\
				& Gall bladder (0.00)& Bone fracture (0.00)& Cancer (0.00)& Colonic stricture (0.00) \\
				& GI bleed -- no var. (0.00)& GI bleed -- var. (0.00)& & \\ \hline
				\multirow{6}{*}{\makecell{Cluster\\7}}& ABPA (0.83)& Diabetes (0.78)& Osteopenia (0.25)& Osteoporosis (0.24)\\
				& Liver Enzymes (0.15)& Intest. Obstruction (0.12)& Liver Disease (0.09)& Hypertension (0.07) \\
				& Hearing Loss (0.07)& Arthropathy (0.06)& Depression (0.04)& Cirrhosis (0.02) \\
				& Asthma (0.01)& Bone fracture (0.01)& Kidney Stones (0.01)& Hemoptysis (0.01) \\
				& Cancer (0.00)& Pancreatitis (0.00)& Gall bladder (0.00)& Colonic stricture (0.00) \\
				& GI bleed -- no var. (0.00)& GI bleed -- var. (0.00)& & \\ \hline
				\multirow{6}{*}{\makecell{Cluster\\8}}& Diabetes (0.94)& Liver Disease (0.83)& Liver Enzymes (0.43)& Osteopenia (0.30) \\
				& Hearing Loss (0.11)& Osteoporosis (0.10)& Intest. Obstruction (0.09)& Cirrhosis (0.08) \\
				& Depression (0.08)& ABPA (0.07)& Arthropathy (0.06)& Hypertension (0.05) \\
				& Kidney Stones (0.05)& Asthma (0.02)& Hemoptysis (0.01)& Bone fracture (0.01) \\
				& Cancer (0.00)& Pancreatitis (0.00)& Gall bladder (0.00)& Colonic stricture (0.00) \\
				& GI bleed -- no var. (0.00)& GI bleed -- var. (0.00)& & \\ \hline
				\multirow{6}{*}{\makecell{Cluster\\9}}& Asthma (0.89)& Osteopenia (0.26)& ABPA (0.19)& Arthropathy (0.14) \\
				& Intest. Obstruction (0.11)& Depression (0.08)& Osteoporosis (0.08)& Diabetes (0.06) \\
				& Liver Enzymes (0.06)& Hemoptysis (0.03)& Hypertension (0.02)& Liver Disease (0.02) \\
				& Pancreatitis (0.02)& Bone fracture (0.01)& Cancer (0.01)& Cirrhosis (0.01) \\
				& Gall bladder (0.01)& Hearing Loss (0.01)& Kidney Stones (0.00)& Colonic stricture (0.00) \\
				& GI bleed -- no var.	(0.00)& GI bleed -- var. (0.00)& & \\ \hline
				\multirow{6}{*}{\makecell{Cluster\\10}}& Osteopenia (0.82)& Diabetes (0.81)& Arthropathy (0.23)& Depression (0.19)\\
				& Liver Enzymes (0.18)& Hypertension (0.16)& Hearing Loss (0.10)& Liver Disease (0.10)\\
				& Osteoporosis (0.10)& Intest. Obstruction (0.09)& ABPA (0.09)& Kidney Stones (0.07)\\
				& Cirrhosis (0.05)& Asthma (0.01)& Cancer (0.00)& GI bleed -- var. (0.00)\\
				& Bone fracture (0.00)& Hemoptysis (0.00)& Pancreatitis (0.00)& Gall bladder (0.00) \\
				& Colonic stricture (0.00)& GI bleed -- no var. (0.00)& & \\ \hline
				\multirow{6}{*}{\makecell{Cluster\\11}}& Osteopenia (0.77)& Liver Enzymes (0.18)& Arthropathy (0.12)& Depression (0.09) \\
				& Hypertension (0.06)& Diabetes (0.06)& Hearing Loss (0.06)& ABPA (0.05) \\
				& Liver Disease (0.05)& Osteoporosis (0.04)& Intest. Obstruction (0.04)& Cirrhosis (0.02) \\
				& Asthma (0.02)& Pancreatitis (0.02)& Bone fracture (0.01)& Cancer (0.01) \\
				& Kidney Stones (0.00)& Gall bladder (0.00)& Colonic stricture (0.00)& Hemoptysis (0.00)\\
				& GI bleed -- no var. (0.00)& GI bleed -- var. (0.00)& & \\
				\hline
			\end{tabular}
		\end{center}
	\end{table*}

	\subsection{How Does the Temporal Phenotypes Change over Time?}
\begin{figure*}[h!]
		\centering
		\includegraphics[width=1.0\linewidth, trim= 0.1 0.1 0.1 0.1]{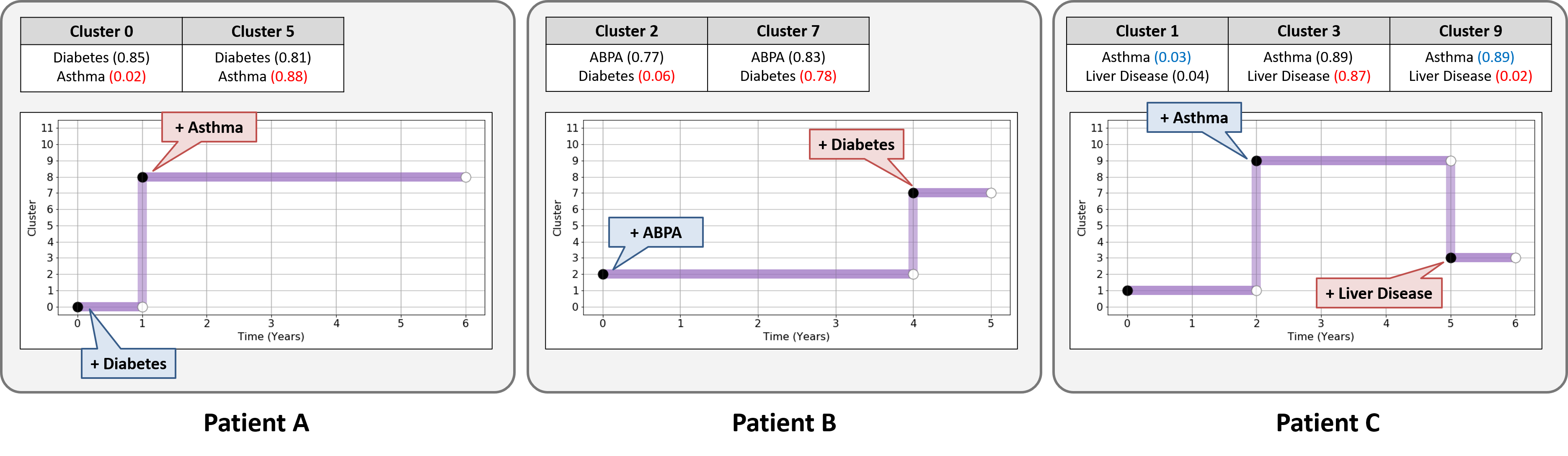}
		\caption{An illustration of run-time examples of \proposed~on three representative patients.}
		\label{fig:cluster_over_time}
	\end{figure*}

	In this subsection, we demonstrate run-time examples of how \proposed~flexibly updates the cluster assignments over time with respect to the future development of comorbidities in the next year. Figure \ref{fig:cluster_over_time} illustrates three representative patients:
	\begin{itemize}[leftmargin=1.5em]
		\item\textbf{ Patient A} had diabetes from the beginning of the study and developed asthma as an additional comorbidity at $t=2$. Accordingly, \proposed~changed the temporal phenotype assigned to this patient from Cluster 0, which consists of patients who are very likely to develop diabetes but very unlikely to develop asthma in the next year, to Cluster 5, which consists of patients who are likely to develop both diabetes and asthma in the next year, at $t=1$.	
		\item \textbf{Patient B} had ABPA from the beginning of the study and developed diabetes at $t=5$. Similarly, \proposed~changed the temporal phenotype assigned to this patient from Cluster 2, which consists of patients who are likely to develop ABPA but not diabetes in the next year, to Cluster 7, which consists of patients who are likely to develop both ABPA and diabetes in the next year, at $t=4$.
		\item \textbf{Patient C} had no comorbidity at the beginning of the study, and developed asthma and liver disease as additional comorbidities, respectively at $t=3$ and $t=6$. \proposed~changed the temporal phenotypes assigned to this patient from Cluster 1 to Cluster 9 at $t=2$ and then to Cluster 3 at $t=5$. The changes in the temporal phenotypes were consistent with the actual development of asthma and liver disease considering the distribution of comorbidity development in the next year -- that is, Cluster 1 consists of patients who are not likely to develop any comorbidities in the next year, Cluster 9 consists of patients who are likely to develop asthma but not liver disease, and Cluster 3 consists of patients who are likely to develop asthma and liver disease in the next year.
	\end{itemize}

	\clearpage	

\begin{thebibliography}{28}
	\providecommand{\natexlab}[1]{#1}
	\providecommand{\url}[1]{\texttt{#1}}
	\expandafter\ifx\csname urlstyle\endcsname\relax
	\providecommand{\doi}[1]{doi: #1}\else
	\providecommand{\doi}{doi: \begingroup \urlstyle{rm}\Url}\fi
	
	\bibitem[Aghabozorgi et~al.(2015)Aghabozorgi, Shirkhorshidi, and
	Wah]{Aghabozorgi:15}
	Aghabozorgi, S., Shirkhorshidi, A.~S., and Wah, T.~Y.
	\newblock Time-series clustering – a decade review.
	\newblock \emph{Information Systems}, 53:\penalty0 16--38, May 2015.
	
	\bibitem[Baytas et~al.(2017)Baytas, Xiao, Zhang, Wang, Jain, and
	Zhou]{Baytas:17}
	Baytas, I.~M., Xiao, C., Zhang, X., Wang, F., Jain, A.~K., and Zhou, J.
	\newblock Patient subtyping via time-aware lstm networks.
	\newblock \emph{In Proceedings of the 23rd ACM SIGKDD Conference on Knowledge
		Discovery and Data Mining (KDD 2017)}, 2017.
	
	\bibitem[Blockeel et~al.(2017)Blockeel, Dzeroski, Struyf, and
	Zenko]{Blockeel:17}
	Blockeel, H., Dzeroski, S., Struyf, J., and Zenko, B.
	\newblock \emph{{Predictive Clustering}}.
	\newblock Springer New York, 2017.
	
	\bibitem[Boudier et~al.(2019)Boudier, Chanoine, Accordini, Anto, {n}a,
	Bousquet, Demoly, Garcia-Aymerich, Gormand, Heinrich, Janson, K\"{u}nzli,
	Matran, Pison, Raherison, Sunyer, Varraso, Jarvis, Leynaert, Pin, and
	Siroux]{Boudier:19}
	Boudier, A., Chanoine, S., Accordini, S., Anto, J.~M., {n}a, X.~B., Bousquet,
	J., Demoly, P., Garcia-Aymerich, J., Gormand, F., Heinrich, J., Janson, C.,
	K\"{u}nzli, N., Matran, R., Pison, C., Raherison, C., Sunyer, J., Varraso,
	R., Jarvis, D., Leynaert, B., Pin, I., and Siroux, V.
	\newblock Data‐driven adult asthma phenotypes based on clinical
	characteristics are associated with asthma outcomes twenty years later.
	\newblock \emph{Allegy}, 74(5):\penalty0 953--963, May 2019.
	
	\bibitem[Fortuin et~al.(2019)Fortuin, H\"{u}ser, Locatello, Strathmann, and
	R\"{a}tsch]{Fortuin:19}
	Fortuin, V., H\"{u}ser, M., Locatello, F., Strathmann, H., and R\"{a}tsch, G.
	\newblock {SOM-VAE}: Interpretable discrete representation learning on time
	series.
	\newblock \emph{In Proceedings of the 7th International Conference on Learning
		Representations (ICLR 2019)}, 2019.
	
	\bibitem[Giannoula et~al.(2018)Giannoula, Gutierrez-Sacrist\'{a}n, Bravo, Sanz,
	and Furlong]{Giannoula:18}
	Giannoula, A., Gutierrez-Sacrist\'{a}n, A., Bravo, A., Sanz, F., and Furlong,
	L.~I.
	\newblock Identifying temporal patterns in patient disease trajectories using
	dynamic ping: A population-based study.
	\newblock \emph{Scientific Reports}, 8(4216):\penalty0 1--14, March 2018.
	
	\bibitem[Glorot \& Bengio(2010)Glorot and Bengio]{Xavier:10}
	Glorot, X. and Bengio, Y.
	\newblock Understanding the difficulty of training deep feedforward neural
	networks.
	\newblock \emph{In Proceedings of the 13th International Conference on
		Artificial Intelligence and Statistics (AISTATS 2010)}, 2010.
	
	\bibitem[Hochreiter \& Schmidhuber(1997)Hochreiter and Schmidhuber]{LSTM:97}
	Hochreiter, S. and Schmidhuber, J.
	\newblock Long short-term memory.
	\newblock \emph{Neural Computation}, 9(8):\penalty0 1735--1780, 1997.
	
	\bibitem[Hubert \& Arabie(1985)Hubert and Arabie]{metric:ARI}
	Hubert, L. and Arabie, P.
	\newblock Comparing partitions.
	\newblock \emph{Journal of Classification}, 2(1):\penalty0 193--218, December
	1985.
	
	\bibitem[Kingma \& Ba(2014)Kingma and Ba]{Adam:14}
	Kingma, D.~P. and Ba, J.
	\newblock Adam: A method for stochastic optimization.
	\newblock \emph{arXiv preprint arXiv:1412.6980}, 2014.
	
	\bibitem[Konda \& Tsitsiklis(2000)Konda and Tsitsiklis]{actor_critic:00}
	Konda, V.~R. and Tsitsiklis, J.~N.
	\newblock Actor-critic algorithms.
	\newblock \emph{In Proceedings of the 13th Conference on Neural Information
		Processing Systems (NIPS 2000)}, 2000.
	
	\bibitem[Lee et~al.(2019)Lee, Yoon, and van~der Schaar]{Changhee:TBME19}
	Lee, C., Yoon, J., and van~der Schaar, M.
	\newblock Dynamic-deephit: A deep learning approach for dynamic survival
	analysis with competing risks based on longitudinal data.
	\newblock \emph{IEEE Transactions on Biomedical Engineering}, April 2019.
	
	\bibitem[Luong \& Chandola(2017)Luong and Chandola]{Luong:17}
	Luong, D. T.~A. and Chandola, V.
	\newblock A k-means approach to clustering disease progressions.
	\newblock \emph{In Proceedings of the 5th IEEE International Conference on
		Healthcare Informatics (ICHI)}, 2017.
	
	\bibitem[Madiraju et~al.(2018)Madiraju, Sadat, Fisher, and
	Karimabadi]{Madiraju:18}
	Madiraju, N.~S., Sadat, S.~M., Fisher, D., and Karimabadi, H.
	\newblock Deep temporal clustering: Fully unsupervised learning of time-domain
	features.
	\newblock \emph{arXiv preprint arXiv:1802.01059}, 2018.
	
	\bibitem[Ramos et~al.(2017)Ramos, Quon, Heltshe, Mayer-Hamblett, Lease, Aitken,
	Weiss, and Goss]{CF_ref:12}
	Ramos, K.~J., Quon, B.~S., Heltshe, S.~L., Mayer-Hamblett, N., Lease, E.~D.,
	Aitken, M.~L., Weiss, N.~S., and Goss, C.~H.
	\newblock Heterogeneity in survival in adult patients with cystic fibrosis with
	{FEV}$_{1}$ $< 30\%$ of predicted in the united states.
	\newblock \emph{Chest}, 151(6):\penalty0 1320--1328, June 2017.
	
	\bibitem[Ranganath et~al.(2016)Ranganath, Perotte, Elhadad, and
	Blei]{Ranganath:16}
	Ranganath, R., Perotte, A., Elhadad, N., and Blei, D.
	\newblock Deep survival analysis.
	\newblock \emph{In Proceedings of the 1st Machine Learning for Healthcare
		Conference (MLHC 2016)}, 2016.
	
	\bibitem[Ratanamahatana et~al.(2005)Ratanamahatana, Keogh, Bagnall, and
	Lonardi]{Ratanamahatana:05}
	Ratanamahatana, C.~A., Keogh, E., Bagnall, A.~J., and Lonardi, S.
	\newblock A novel bit level time series representation with implications for
	similarity search and clustering.
	\newblock \emph{In Proceedings of the 9th Pacific-Asia Conference on Knowledge
		Discovery and Data Mining (PAKDD 2005)}, 2005.
	
	\bibitem[Ronan et~al.(2017)Ronan, Elborn, and Plant]{comorbidity_ref:2}
	Ronan, N.~J., Elborn, J., and Plant, B.~J.
	\newblock Current and emerging comorbidities in cystic fibrosis.
	\newblock \emph{Presse Med.}, 46(6):\penalty0 125--138, June 2017.
	
	\bibitem[Rousseeuw(1987)]{Silhouette_Index}
	Rousseeuw, P.~J.
	\newblock Silhouettes: a graphical aid to the interpretation and validation of
	cluster analysis.
	\newblock \emph{Computational and Applied Mathematics}, 20:\penalty0 53--65,
	1987.
	
	\bibitem[Rusanov et~al.(2016)Rusanov, Prado, and Weng]{Rusanov:16}
	Rusanov, A., Prado, P.~V., and Weng, C.
	\newblock Unsupervised time-series clustering over lab data for automatic
	identification of uncontrolled diabetes.
	\newblock \emph{In Proceedings of the 4th IEEE International Conference on
		Healthcare Informatics (ICHI)}, 2016.
	
	\bibitem[Samal et~al.(2011)Samal, Wright, Wong, Linder, and Bates]{Samal:11}
	Samal, L., Wright, A., Wong, B., Linder, J., and Bates, D.
	\newblock Leveraging electronic health records to support chronic disease
	management: the need for temporal data views.
	\newblock \emph{Informatics in Primary Care}, 19(2):\penalty0 65--74, 2011.
	
	\bibitem[van~den Oord et~al.(2017)van~den Oord, Vinyals, and
	Kavukcuoglu]{VQ-VAE:17}
	van~den Oord, A., Vinyals, O., and Kavukcuoglu, K.
	\newblock Neural discrete representation learning.
	\newblock \emph{In Proceedings of the 31st Conference on Neural Information
		Processing Systems (NIPS 2017)}, 2017.
	
	\bibitem[Vinh et~al.(2010)Vinh, Epps, and Bailey]{metric:NMI}
	Vinh, N.~X., Epps, J., and Bailey, J.
	\newblock Information theoretic measures for clusterings comparison: Variants,
	properties, normalization and correction for chance.
	\newblock \emph{Journal of Machine Learning Research}, 11(1):\penalty0
	2837--2854, October 2010.
	
	\bibitem[Wami et~al.(2013)Wami, Buntinx, Bartholomeeusen, Goderis, Mathieu, and
	Aerts]{comorbidity_ref:1}
	Wami, W.~M., Buntinx, F., Bartholomeeusen, S., Goderis, G., Mathieu, C., and
	Aerts, M.
	\newblock Influence of chronic comorbidity and medication on the efficacy of
	treatment in patients with diabetes in general practice.
	\newblock \emph{The British Journal of General Practice}, 63(609):\penalty0
	267--273, March 2013.
	
	\bibitem[Xie et~al.(2017)Xie, Girshick, and Farhadi]{Xie:16}
	Xie, J., Girshick, R., and Farhadi, A.
	\newblock Unsupervised deep embedding for clustering analysis.
	\newblock \emph{In Proceedings of the 33rd International Conference on Machine
		Learning (ICML 2016)}, 2017.
	
	\bibitem[Yang et~al.(2017)Yang, Fu, Sidiropoulos, and Hong]{Yang:17}
	Yang, B., Fu, X., Sidiropoulos, N.~D., and Hong, M.
	\newblock Towards k-means-friendly spaces: Simultaneous deep learning and
	clustering.
	\newblock \emph{In Proceedings of the 34th International Conference on Machine
		Learning (ICML 2017)}, 2017.
	
	\bibitem[Yoon et~al.(2017)Yoon, Davtyan, and van~der Schaar]{Jinsung:JBHI17}
	Yoon, J., Davtyan, C., and van~der Schaar, M.
	\newblock Discovery and clinical decision support for personalized healthcare.
	\newblock \emph{IEEE J Biomed Health Inform.}, 21(4):\penalty0 1133--1145,
	2017.
	
	\bibitem[Zhang et~al.(2019)Zhang, Chou, Liang, Xiao, Zhao, Sarva, Henchcliffe,
	and Wang]{FeiWang:18}
	Zhang, X., Chou, J., Liang, J., Xiao, C., Zhao, Y., Sarva, H., Henchcliffe, C.,
	and Wang, F.
	\newblock Data-driven subtyping of parkinson’s disease using longitudinal
	clinical records: A cohort study.
	\newblock \emph{Scientific Reports}, 9(797):\penalty0 1--12, January 2019.
	
\end{thebibliography}

\clearpage
\begin{thebibliography}{3}
	\providecommand{\natexlab}[1]{#1}
	\providecommand{\url}[1]{\texttt{#1}}
	\expandafter\ifx\csname urlstyle\endcsname\relax
	\providecommand{\doi}[1]{doi: #1}\else
	\providecommand{\doi}{doi: \begingroup \urlstyle{rm}\Url}\fi
	
	\bibitem[Fortuin et~al.(2019)Fortuin, H\"{u}ser, Locatello, Strathmann, and
	R\"{a}tsch]{Fortuin:19}
	Fortuin, V., H\"{u}ser, M., Locatello, F., Strathmann, H., and R\"{a}tsch, G.
	\newblock {SOM-VAE}: Interpretable discrete representation learning on time
	series.
	\newblock \emph{In Proceedings of the 7th International Conference on Learning
		Representations (ICLR 2019)}, 2019.
	
	\bibitem[Konda \& Tsitsiklis(2000)Konda and Tsitsiklis]{actor_critic:00}
	Konda, V.~R. and Tsitsiklis, J.~N.
	\newblock Actor-critic algorithms.
	\newblock \emph{In Proceedings of the 13th Conference on Neural Information
		Processing Systems (NIPS 2000)}, 2000.
	
	\bibitem[Yang et~al.(2017)Yang, Fu, Sidiropoulos, and Hong]{Yang:17}
	Yang, B., Fu, X., Sidiropoulos, N.~D., and Hong, M.
	\newblock Towards k-means-friendly spaces: Simultaneous deep learning and
	clustering.
	\newblock \emph{In Proceedings of the 34th International Conference on Machine
		Learning (ICML 2017)}, 2017.
	
\end{thebibliography}
\end{document}